# Analytic insights into nonlocal energy transport: Steady State Fokker Planck theory in arbitrary Z plasmas

Wallace Manheimer
1108 Whitney Lane
Allendale, NJ 07401
wallymanheimer@yahoo.com
Consultant to the Naval Research Laboratory

## Abstract

The generation of energetic electrons in laser fusion in an important issue.  The electrons may either arise from a laser plasma instability, or from the uncoupled high temperature tail of a Maxwellian distribution.  To study these in a laser fusion context, it is important to find a method accurate enough to be useful, and simple enough to be incorporated into a radiation hydrodynamics numerical simulation, the main workhorse for studying the laser fusion target.   That is why analytic insights become important, they allow one to simplify the Fokker Planck theory so that a solution of it can be incorporated into a radiation hydrodynamic (rad-hydro) simulation. This work develops and analyzes a steady state Fokker Planck theory for plasmas of arbitrary Z.  It developes a method of solving the simplified Fokker Planck method with a technique called sparse eigenfunction analysis.  This method appears to work reasonably well when compared to the experimental results from the Rochester/NIF on plastic spherical targets with and without a silicon layer.

I. Introduction

This work extends earlier work on a steady state Fokker Planck model for the deposition of energetic electrons into a laser fusion target. These earlier theories were for Z=1 plasmas in planar (1,2) and spherical (3) configurations. This work extends that theory to plasmas of arbitrary arbitrary Z, and even to structured targets with several different regions having different Z's. What it cannot model yet, is a plasma with strongly varying Z as a function of position, such as a gold plasma expanding from an intense laser radiation of a hohlraum wall.

In a laser fusion target, energetic electrons could have several important effects. First of all, most of their energy is deposited in the ablating shell, which could alter the shell structure as the NIF/URLLE (University of Rochester Laboratory for Laser Energetics) experiments showed (4). It could broaden the shell, and this might have a stabilizing effect on fluid instabilities like the Rayleigh Taylor instability. However the broadened shell might not be as effective in compressing the fuel. Also, some of the electrons are deposited not only in the shell, but also in the fuel, causing it to preheat, making fusion more difficult. The energy put into energetic electrons is not necessarily the energy deposited in the fuel (3,5) . As Christopherson et al (5) put it "What is critically important is not the total amount of energy transferred to hot electrons, but the amount of that energy that such electrons deposit in the DT fuel…." It does not take much fuel preheat to cause the implosion to fail. Hence the presence or absence of energetic electrons could have either a positive or negative effect on laser fusion, but more likely negative. The study of them is undoubtedly important for its development. This work does study the DT fuel preheat, and indeed shows that it is a tiny fraction of the total preheat generated by the energetic electrons, at least in the examples considered here.

Since the principle way in which laser fusion targets are studied is with radiation hydrodynamic (rad-hydro) fluid simulations, it is important that whatever theory is used for energetic electrons, fits in well with the rad hydro architecture and does not significantly increase the simulation running time. Hence the theory developed here cannot be as accurate as a particle in cell or Monte Carlo simulation. These simulations for a single realistic configuration (i.e a single instant of the implosion), take about as long to do as the entire hydro simulation, so they can hardly be done economically at each fluid time step. The theory developed here seems to be able to clear this particular hurdle.

These energetic electrons have at least two possible sources, first a laser plasma instability (6,7), and second the high energy tail of a Maxwellian. These electrons' long mean free path may decouple them from the thermal plasma where they were produced. This paper concentrates on the former, a future work will discuss the latter. One reason to concentrate on the former, and something which motivates this work, is a recent series of experiments which the scientists of the University of Rochester Laboratory for Laser Energetics (URLLE) performed on planar and on structured spherical targets at the National Ignition Facility (NIF) at the Lawrence Livermore National Laboratory (LLNL) (4, 8,). Using the hard X-ray spectrum they can deduce the temperature of the hot electrons in the NIF/URLLE experiments, as well as in earlier experiments on their OMEGA facility (9). So far, based admittedly on the author's incomplete knowledge of the experimental parameters, and so far, on the fact that the theory has not yet been incorporated into a rad hydro code, it looks like the theory here gives a reasonable model for the experiment.

Furthermore, URLLE performed experiments to measure the fuel preheat on their own OMEGA laser (5). They performed these experiments by irradiating two mostly plastic, equivalent mass spheres, one with the deuterium fuel, one without it, and measuring hard X-ray emission in each. In the sphere without the fuel, the deuterium mass is replaced with plastic mass, which radiates more due to its higher Z. Measuring the difference in the two radiation signatures gives a good indication of the energetic electron energy deposited in the fuel. The theory developed here also can predict the fuel preheat, and we apply it to the NIF/URLLE experiment.

Section II discusses the instabilities which the URLLE group have seen experimentally and have performed various measurements on. Section III discusses the Fokker Planck equation. It uses a two term Legendre expansion of the angular dependence, and various other approximations made to give analytic solutions relevant to incorporation in a rad hydro code. Section IV discusses some preliminary aspects, particularly the relation between the steady state Fokker Planck theory and what has been called the two fluid diffusion model (10). Section V develops one possible way of solving the reduced equation, a method the author has called the 'sparse eigenfunction' method. It was originally developed in (3) and here is extended to higher Z. Section VI formulates the solution for the spatial dependence of the energetic electron energy flux for a target like those used in the URLLE/NIF experiments. Section VII examines the effect of higher order terms in the Legendre expansion. For Z=1, the 2 term approximation is at best only qualitatively correct; for Z=2 and 3, the approximation is reasonably valid, and for larger Z's, the two term expansion is quite good, at least for the targets examined here. Section VIII discusses the spherical corrections, and Section IX discusses the theoretical results as may be applicable to the URLLE/NIF measurements of spatial deposition of energetic electron energy. Section X draws conclusions. The Appendix introduces another way of finding approximate solution, which while not valid everywhere, is a quick way of getting a first approximation to the to the energy flux via a simple formula.

II. The Instabilities

The two electron instabilities which have been studied experimentally, especially at the University of Rochester Laboratory for Laser Energetics (URLLE), have been the two plasmon decay instability (or $2\omega_p$), and the absolute Raman scattering instability (SRS) (6,7). They did experiments on both their OMEGA ($\Omega$) laser facility, and the NIF laser at the Lawrence Livermore National Laboratory (LLNL).

Since the NIF creates both hotter and larger plasmas, they found that on $\Omega$, they typically saw the $2\omega_p$ instability; whereas on NIF, SRS. Their published values for instability threshold are (6):

SRS: $I > 2377/L_n^{4/3}$ and $2\omega_p$: $I > 233 T_e/L_n$

Here I is the laser irradiance in units of $10^{14}$ W/cm$^2$, $L_n$ is the density gradient scale length in microns, and $T_e$ is the electron temperature in keV.

The URLLE has done a long series of experiments and theoretical studies of energetic electron production by both the $2\omega_p$ instability (2PD) and the absolute stimulated Raman scattering instability (SRS). Their experiments on the OMEGA laser showed that the 2PD instability was the dominant effect. On the other hand, their experiments on the NIF laser, which produced much hotter and larger plasmas, due to the much higher laser energy showed that the SRS instability was the dominant effect. Both of these experiments were performed on planar targets.

It is interesting that the former showed a strong dependence of hot electron temperature on laser irradiance, or more precisely on the ratio of laser irradiance to threshold irradiance. However on the NIF experiments, the hot electron temperature was pretty much the same independent of irradiance as long it was above threshold. This is shown in Figure 1.

A

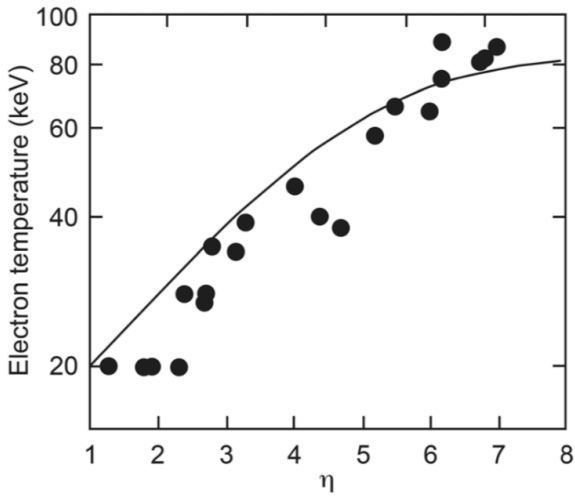

B

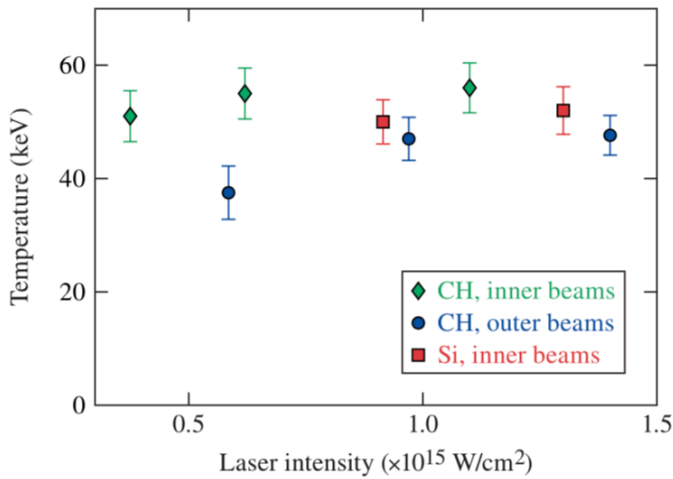

Figure (1): A: The hot electron temperature from 2012 (9) from OMEGA from the $2\omega_p$ instability with planar targets. $\eta$ is the ratio of laser irradiance to threshold irradiance  B: The hot electron temperature 2020 (2) NIF experiments on planar targets.

The results of the NIF experiments seem to show that the hot electron temperature depends mostly on the spectrum of the electrostatic waves produced, whereas in the OMEGA experiments, it depends somehow on the irradiance.

More recently their experiments have shifted to spherical targets.
The spherical targets here were mostly plastic shells, but with various impurities in the plastic, and with deuterium inside. These targets are all mass equivalent and had the same outer layer, so that the SRS signature from all of them was the same. The plastic often had a high Z impurity, for instance copper or germanium, which radiated more, and these were placed at various positions in the plastic, and with various thicknesses. These impurities were placed deep enough in the plastic that in the time of the experiment, they were not present in the blowoff plasma. Deconvolving the radiation, and using Monte Carlo simulations, they could get a good idea of the spatial profile of the energetic electrons in the ablated and non ablated plastic.

In another series of experiments, they put in a thin silicon layer nearer to the edge of the shell. The purpose here was to steepen the gradient near the quarter critical density so as to partially stabilize the SRS instability. Their rad-hydro simulations did in fact predict that the gradient would be greater at the quarter critical density. Specifically the experiments motivating us are shown below (taken from 4 ):

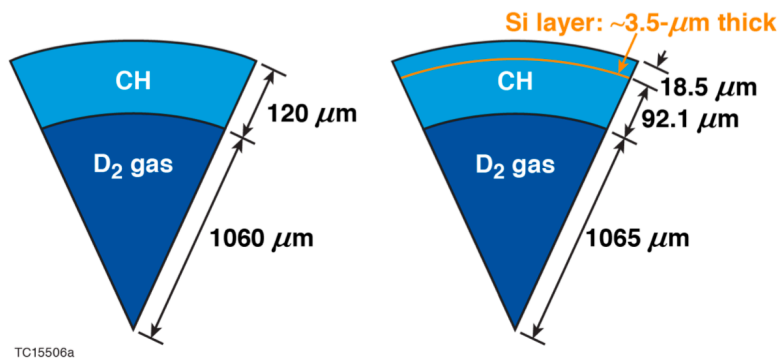

Figure (2): Two of the imploded plastic shells used in the NIF/URLLE .

In each of the experiments, the laser pulse is shown below in Figure 3. The laser producing this pulse is a frequency tripled Nd laser with a wavelength of 351 nanometers (nm).

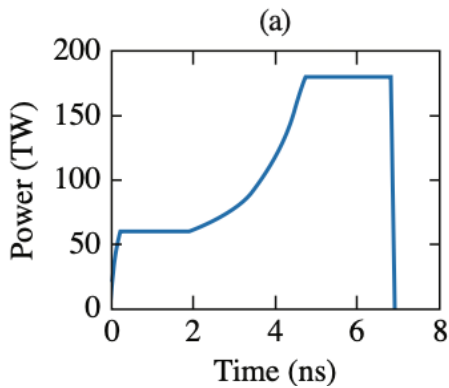

Figure (3): The laser pulse from the URLLE/NIF experiments with and without the silicon impurity layer (i.e. 180 TW);

An important aspect of these experiments is that, except for the deuterium, the Z's are greater than one, 5.3 for the fully ionized plastic, and 14 for the fully ionized silicon. Hence this work extends our earlier work to layered targets with a variety of Z's.

The fraction of laser energy put into energetic electrons increases with laser irradiance, as measured by the energetic X-ray energy. The silicon layer reduced the energy transferred to hot electrons, and they interpreted this as an increase in the density gradient and collisionality at the quarter critical density, so that the instability is closer to threshold. In Figure (4) is a plot of URLLE/NIF results of the fraction of laser energy put into fast electrons, as measured by the energy in the hard X-rays, taken from (4).

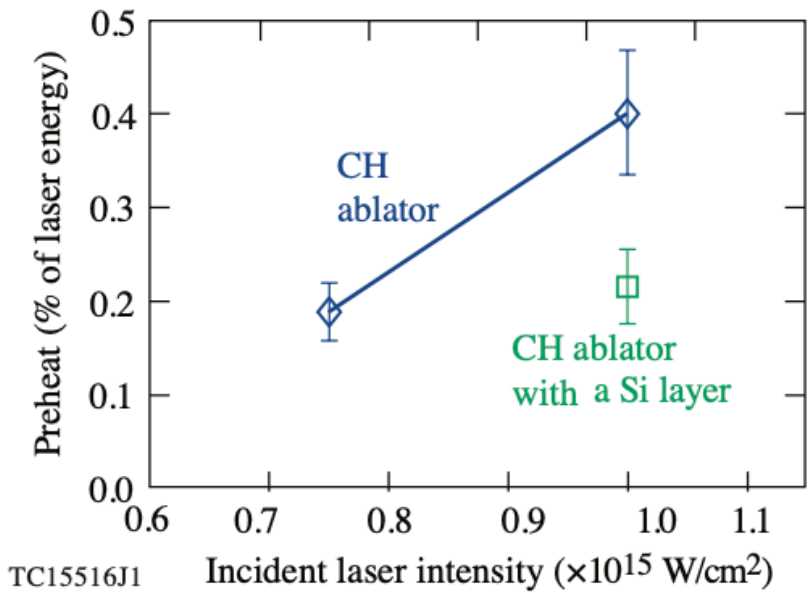

Figure (4): The fraction of laser energy going into hot electrons as a function of irradiance for the plastic ablator, and the plastic with a silicon layer. The silicon layer reduces the preheat roughly equivalent to reducing the irradiance by ~ 25%;

The conclusion seems to be that the hot electron temperature is determined by the phase velocity of the electron plasma wave generated by the instability, and not by the irradiance, but the fraction of laser energy put into energetic electrons increases with laser irradiance. However as Ref (4) points out: "To match the HXR spectrum for all mass equivalent implosions, a fraction of the HXR emission should be generated by hot electrons recirculating in the outer CH corona. This fraction represents 34% plus or minus 10%, to 10% plus or minus 10%....." In other words, they concluded that there are other energetic electrons which never make it into the ablating plasma, but skirt around the outside. They have a difficult time estimating the population of these electrons. We will discuss this further in Appendix B.

The URLLE group made a considerable effort, both in its $\Omega$ experiments (11), and its NIF experiments (8) to show that the electrons are produced in a rather wide conical angle, about 45 degrees. However, this author wonders if the divergence angle is really a very important measurement as regarding the application to laser fusion. This will be discussed in later sections of this paper and in Appendix C.

In addition to the experiments, there have also been particle simulations of stimulated Raman side scatter. This author participated in one of the first about half a century ago (12). However much more complete, three-dimensional simulations have been performed by Xiao et al (13). Although their parameters do not match very well the URLLE/NIF experiments, they do add considerable insight.

In the simulations, Xiao et al used a 1 micron laser with an irradiance of $1.4 \times 10^{16}$ W/cm$^2$ and an electron temperature of 1 keV. Their plasma was a rectangular box 400 microns long with a density increasing linearly from 0 to $0.3 n_{cr}$ over that range. A plot of their average distribution function $f(p_x)$ over various regions, is shown in Fig (5) (x is the axis along the length of the box).

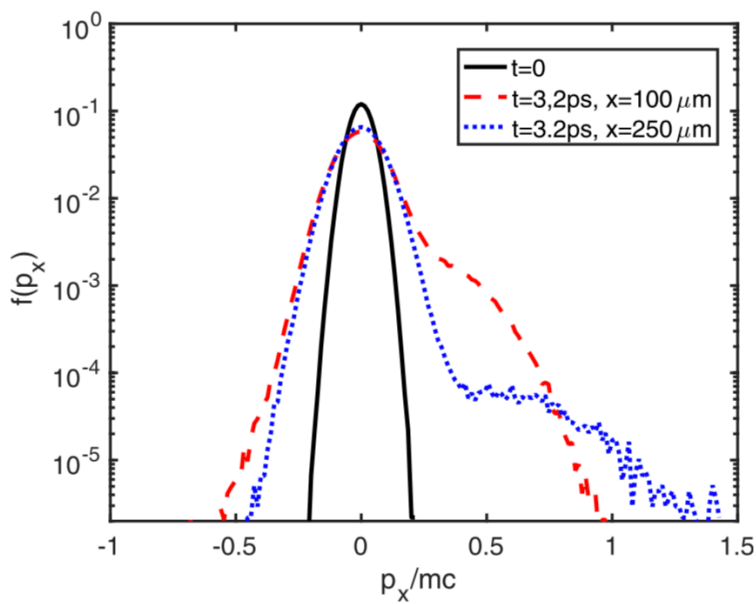

Figure (5): The distribution function at various positions, taken from Ref (14). At t=0 it is a Maxwellian with temperature 1 keV. Clearly, as the position approached the high density edge, a separate hot electron plasma is generated.

The hot electron temperature is about 40-50 keV. This is in basic agreement with that measured in the URLLE/NIF experiment even though the plasma and laser irradiance and wavelength are very different from what the NIF experiments produced. This is further evidence that the hot electron temperature is determined principally by the phase velocity of the electron plasma wave produced, and depends only weakly on the laser irradiance and the thermal plasma temperature. Dr. Xiao provided a two-dimensional plot of the electron phase space shown in Fig (6).

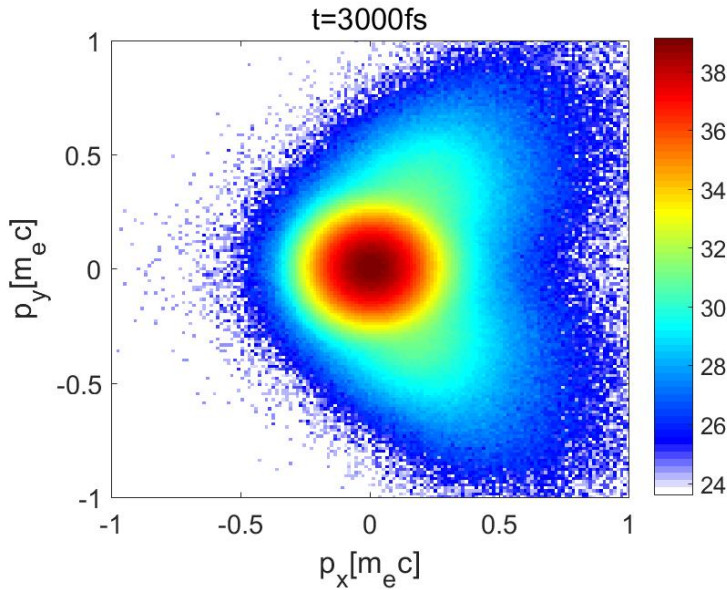

Figure (6): A two dimensional plot of the electron phase space (vertical scale is a log scale) supplied by Dr Xiao for his RSSI simulations, for which the author thanks him.

Clearly this supports the URLLE assertion that the energetic electrons are produced in a cone of about 45 degrees about the direction of laser propagation.

III Fokker Planck Formulation

The idea is to treat the small population of energetic electrons with a Fokker Planck theory. They interact with the electrons and ions of the much denser background plasma, but not with each other. The physical processes which the Fokker Planck model describes are:

1. Angular scattering off the plasma electrons.
2. Angular scattering off the ions. For ions of equivalent charge state Z, this is Z times the scattering off the electrons. For a multi species plasmas, the equivalent Z is

$$Z = \frac{\sum_i n_i Z_i^2}{\sum_i n_i Z_i}$$

Here i is the species index.

3. Dynamical friction on the plasma electrons.
4. Energy diffusion from collisions with the plasma electrons

The last process, energy diffusion, is smaller than the others enumerated by about a factor of T/E where T is the energy of the background plasma, ~ a few kilovolts, and E is the energetic electron energy, ~ 40-50 keV. Hence we neglect energy diffusion of the energetic electrons.

Furthermore, since the target has an electric field which gives an electrostatic potential of order the electron temperature, we neglect the effect of the electric field on the much more energetic electrons.

Let us digress by considering the effect of the dynamical friction of the energetic electron with the more numerous electrons in the main plasma. Consider an energetic particle of velocity $V_o$. The key here is that the slowing down rate greatly increases as the particle slows down with a rate $V^{-2}$. Consider the simplest equation of motion of a particle with initial velocity $V_o$

$$V \frac{dV}{dx} = -\frac{v(V_o)}{V^2} V_o^3 \qquad (1)$$

It is a simple matter to solve and find

$$V = \left[ V_o^4 - 4v(V_o) V_o^3 x \right]^{1/4} \qquad (2)$$

In other words, the particle cannot even travel a single mean free path as defined at its initial energy; it can travel only a quarter of this mean free path. This is the reason that the steady state Fokker Planck model gives much less fuel preheat than the many Krook models that have been published. Those Krook models have no dynamical friction in them, but simply reduce the flux by a factor of e every mean free path. Hence even after several e foldings, there are many more energetic electrons that make it into the fuel and provided much more preheat (1,2).

The appropriate Fokker Planck equation for the distribution function of the energetic electrons, f in spherical coordinates is

$$\frac{\partial f}{\partial t} + V\cos\theta \frac{\partial f}{\partial r} - \frac{V\sin\theta}{r}\frac{\partial f}{\partial \theta} = \frac{1}{V^2}\frac{\partial}{\partial V}V^3 \nu_e(V) f + \frac{\nu_e}{2}(1+Z)\left[\frac{1}{\sin\theta}\frac{\partial}{\partial \theta}\sin\theta\frac{\partial}{\partial \theta} + \frac{1}{\sin^2\theta}\frac{\partial^2}{\partial \phi^2}\right]f$$

(3)

Here the energetic electrons are colliding only with a much denser, cooler background thermal plasma. The first term on the right hand side describes the dynamical friction of the fast electrons as they collide with the background electrons. The second term describes their angular scattering from collisions with both the thermal electrons and ions. Since the energetic electron energy is much greater than the plasm temperature, we neglect the energy diffusion of the energetic electrons. The two angles shown on the right hand side are the polar angles of the velocity vector with respect to the radius vector. Thus these angles depend on position as the unit vector in the radial direction depends on position. The third term on the left hand side is the spherical correction to the planar theory (3). The quantity $\nu_e$ is the fundamental electron collision frequency

$$\nu_e = \frac{4\pi n e^4}{m^2 V^3}\Lambda \qquad \nu_e = 3.9 \times 10^{-6}\frac{n}{\Xi^{3/2}}\Lambda$$

in cgs or (4)

where n is the electron density in cm$^{-3}$ and $\Xi$ is the electron energy in eV and $\Lambda$ is the Coulomb logarithm. The Coulomb logarithm for the energetic electrons is not the same as that for the thermal electrons, but is considerably larger even in the high density, low temperature plasma where the Coulomb logarithm for the thermal plasma is much lower, as we will discuss shortly.

The approach then is to assume that f has no dependence on ϕ, and expand f in a series of Legendre polynomials $P_i(\theta)$, keeping only the first two terms. That is we take

$$f(V, \theta, r) = \sum_{i=0}^{n} f_i(V, r) P_i(\theta) \tag{5}$$

and take only the first two terms of the expansion. The first two Legendre polynomials are 1 and cosθ. This is a standard approximation, which we will discuss further in a later section. The Fokker Planck equation then breaks into two coupled equations for $f_o$ and $f_1$:

$$\cancel{\frac{\partial f_0}{\partial t}} + \frac{V}{3}\frac{\partial f_1}{\partial r} + \frac{2V}{3r}\cancel{f_1} = \frac{1}{V^2}\frac{\partial}{\partial V} V^3 v_e(x,V) f_0$$

$$\cancel{\cancel{\frac{\partial f_1}{\partial t}}} + V\frac{\partial f_0}{\partial r} = \frac{1}{V^2}\frac{\partial}{\partial V} V^3 \cancel{v\,(x,V) f_1} - v_e(x,V)(1+Z) f_1 \tag{6}$$

The equations as written have various terms struck out, in red or green, depending on the particular approximation used to simplify the equations. Those struck out in green constitute what is called the diffusion approximation (10). Making this approximation, the second equation is a simple algebraic equation for $f_1$. Inserting it in the first equation gives a single equation for $f_o$. In the spatial domain, it is a diffusion equation, in the velocity domain, it describes dynamical friction. It is apparent from the diffusion constant, that length scales as $(Z+1)^{-1/2}$ in the diffusion approximation. Given $f_o$, the heating of the main plasma can easily be calculated from the collisional energy transfer of the two plasma components.

To this author, the diffusion approximation seems difficult to justify.

After all, why neglect the time dependence and dynamical friction in the second equation, but not the first? Furthermore, why neglect the spherical correction (although much of what we will do here also makes this assumption, but only for simplicity, it can be included relatively easily)?

The other approximation, neglecting the terms struck out in red, is simply a steady state approximation. This depends only on the fact that the collision time for the energetic electrons is much less than the fluid time. For a 40 keV electron in a plasma with an electron number density of ~ $10^{22}$ cm$^{-3}$ and a Coulomb logarithm of 8, the electron electron collision time is about 20 picoseconds, and the electron ion time, less, depending on the Z. However the fluid time for the implosion is ~ 10ns for a typical laser fusion implosion. Hence the steady state approximation does seem much easier to justify. As we will see, unlike the diffusion approximation, there is no simple scaling of length with Z+1. However lengths do shrink as Z increases.

The diffusion equation gives $f_o$ for the energetic particles. The background plasma heating is given by the collisional exchange between these two plasmas. In the steady state approximation, $f_1$ is the more natural function to solve for, and the background heating is determined by the divergence of the energetic electron energy flux.

Henceforth we consider only the steady state approximation. We simplify the equation for $f_1$ by making the following transformation of the independent variables.

$$dz = \sqrt{3}\frac{2.7 \times 10^{-13} n(r) \Lambda}{T_e^2} dr \equiv k(r, T_e) dr \qquad \text{and} \qquad w = \varsigma^2 \equiv \left[\frac{\Xi}{T_e}\right]^2 \qquad (7)$$

The independent variable z, which has no Z dependence, is basically the length measured in units of electron-electron collisional mean free path. However, because of the strong increase of the dynamical friction with decreasing electron energy (it goes as $\Xi^{-3/2}$), the electron can only travel about one quarter of its mean free path defined at its initial energy. Also, one must calculate z(r).

Hence to proceed, one must have n(r).  For the URLLE experiment, n(r) is, for our purposes, basically the same for both the experiment with and without the silicon layer.   However, for the former case, much of the blowoff plasma is silicon, presumably with a Z of 14.  Without the silicon, it is plastic (i.e CH) with a Z of 5.3.  The spatial dependence of electron density for the experiment with the silicon, as calculated by URLLE with their LILAC code at 6 ns, is given by Figure (7A), provided by Andrey Solodov of URLLE.  The regions of D, CH and Si are marked.  For the other experiment without the silicon, the Si part is replaced with CH.  Figure (7B) shows z(r).  For completeness, we also show their calculated electron temperature in C.

A

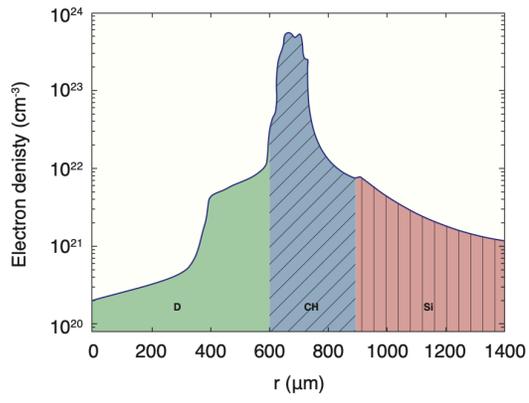

B

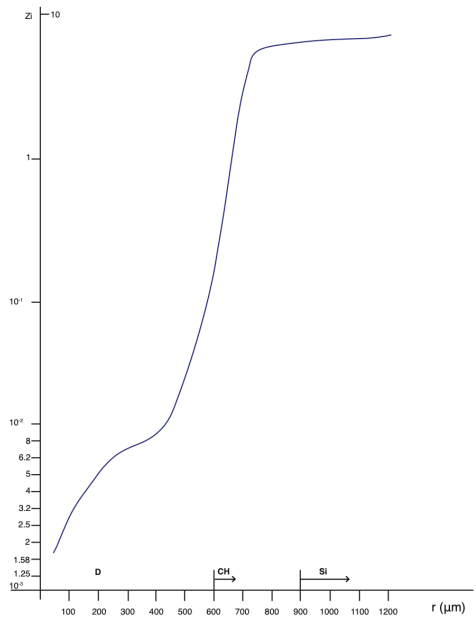

C

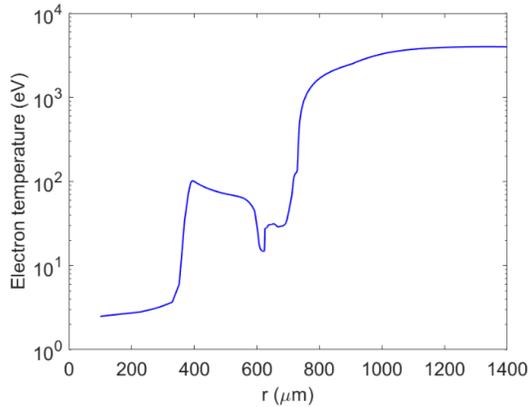

Figure (7) A: The electron density for the URLLE silicon spherical experiment, as calculated by their code LILAC at a time of 6 ns. The regions of deuterium, in green; plastic, blue, slanted lines; and Si, pink, vertical lines  B: z(r) for this case. C: the electron temperature. Provided by A. Solodov of URLLE, for which the author thanks him..

Actually, there are differences in the density profile for the two cases. They involve the density gradient at the quarter critical surface. It is larger for the plasma with the silicon impurity, meaning that the instability is closer to threshold. That, and the higher Z of the silicon plasma give rise to fewer energetic electrons. However, for our purposes, where the distribution function of the energetic electrons is assumed specified at the quarter critical surface, the density profiles can be regarded as the same. Also, the temperature profiles with and without the silicon are basically the same, but with slight differences.

As a comparison, we show the analogous graphs for the NRL simulation of a shock ignition laser implosion used in (2,3, 14) in Figure ( 8 A, B, and C):

A

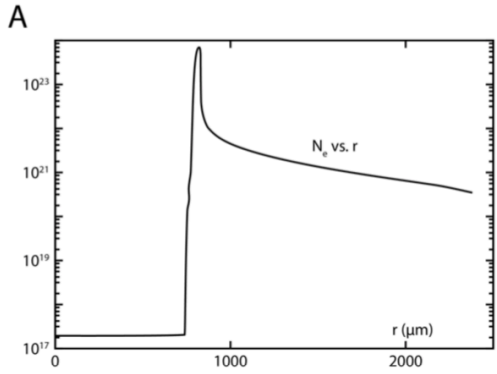

B

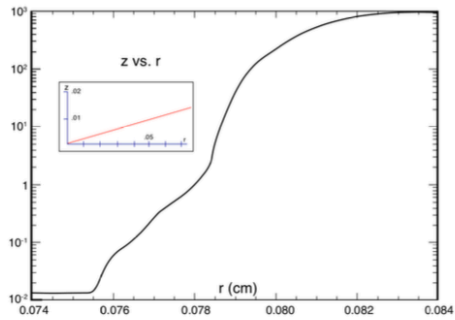

C

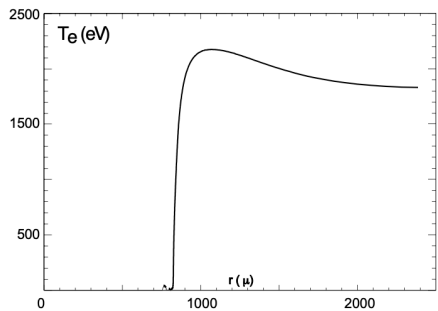

Figure (8): Analogous to Fig (7), but for the NRL shock ignition simulation as calculated by the rad hydro code FAST at a time of 15 ns, just before the final shock is launched, taken from Ref 14. The number density is in $cm^{-3}$, and z is dimensionless. There was assumed to be no separate hot fluid; the theory involved the nonthermal tail of the 2 keV

plasma, the thermal temperature defining, and hence producing, much larger z's.

For energetic electrons in a Dense plasma, the Coulomb logarithm Λ is not necessarily given by the local value, since the distance of closest approach, h/p is determined by the energetic electrons. Also in many cases, there are not enough electrons in a Debye sphere to shield its charge. Hence the interparticle separation can by used for the maximum impact parameter instead. A more accurate expression is given by Eq. (26) of (14). Figure (9) gives the calculated Λ for the case shown in Fig 8, from (14), showing both the classical and modified values Λ for the energetic electrons (here from the tail of the distribution function).

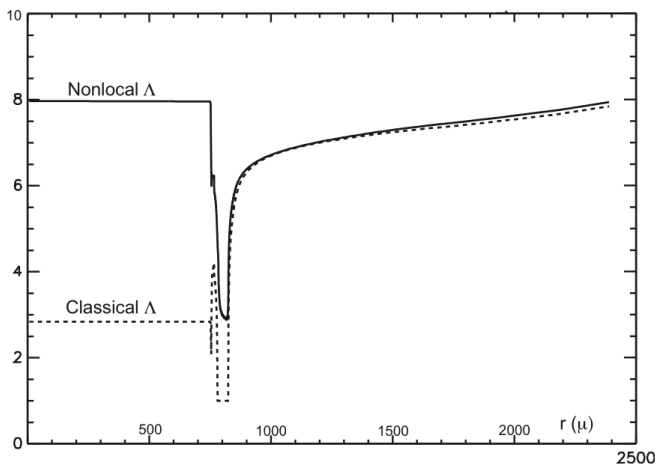

Figure 9: The classical value of L and its modified value accounting for the energetic electrons propagating into the dense plasma,

Notice that in the NRL run, the fuel density is uniform (i.e. r<750), where as in the LILAC runs, there is considerable variation in the fuel density (i.e. r<600). Also the temperature in the NRL run is much smoother than the LILAC run. Notice also that the thermal temperature in the blowoff plasma is about a kilovolt higher in the URLLE/NIF run than in the NRL calculation. Perhaps that is because the NRL simulation used a shorter wavelength laser, 0.25μm as opposed to 0.35 used at NIF. The LILAC run was for an existing experiment, and no effort was made to taylor the target or laser pulse; there is a rather simple rise and a flat top pulse as shown in Fig (3). Hence there are more likely to be waves and other complex phenomena bouncing around.  The FAST result shown is a result of an optimization, optimizing both the target and laser pulse, for maximum computed gain.  For this laser pulse, the fuel simply heats adiabatically as it compresses and maintains uniformity.  Also note that the fuel density is much less in the FAST simulation than for the URLLE/NIF experiment.

The steady state Fokker Planck equation can be combined into a single equation for $f_1$

$$\frac{\partial^2 f_1}{\partial z^2} + 2\frac{\partial}{\partial z}\frac{f_1}{r(z)\left[\frac{dz}{dr}\right]} = \frac{\partial^2 f_1}{\partial w^2} - \frac{1+Z}{4w}\frac{\partial f_1}{\partial w} + \frac{1+Z}{4w^2}f_1$$

(8)

Ultimately, the goal could be to numerically solve this two dimensional partial differential equation in (z,w) space.  An advantage of this approach is that Z can be an arbitrary function of z.  The boundary conditions would be $f_1(z_o,w)$ is specified at $z_o$, the position of the quarter critical surface;  $f_1(0,w) = 0$, that is there is no energy flux into the origin at any w; $f_1(z, \text{infinity}) = 0$;  and $f_1(z,0) = 0$, that is the first order term must be an odd function of w for all z.  Where the boundary conditions are specified on a closed surface in (w,z) space, and the equation is a second order partial differential in that space, the  solutions are stable. That is a small perturbation to the chosen boundary condition gives rise to only a small perturbation to the solution throughout the region.  In Section (V) we develop an approximate solution based on an approximate specification for $f_1(z_o,w)$.   The conditions on the other boundaries are known exactly.

In this paper, we will take a simpler approach and use a particular one dimensional approximation which we develop here, which satisfies the boundary conditions just stated, but one which assumes the target is segmented into a small number of regions, each one with a spatially constant Z, as in the URLLE/NIF experiment.  This approach can easily be incorporated into an operation that a Rad Hydro simulation like LILAC or FAST can implement at each time step.

IV. Some preliminaries

Before embarking on a solution of the steady state Fokker Planck equation to determine the deposition of the energetic electrons, we first consider some preliminaries.

First of all, what is required for modeling the nonthermal energetic electrons in a fluid simulation is their energy flux. The URLLE/NIF as well as the earlier URLLE OMEGA experiments did not measure this. What they did measure is the hard X-ray temperature and assumed that this was the temperature of the energetic electron subpopulation. Then, since their laser pulses were typically flat top (perhaps with a prepulse) and of shorter duration than a full spherical implosion would have been, they measure the total energy of the energetic electrons by knowing the total hard X-ray energy.

This is not a viable approach to simulating a laser fusion implosion, as the laser pulse is much longer and has much more temporal variation than do the pulses used in the experiments. The basic assumption we make here is that the ratio of total energetic electron energy to total laser energy, is equal to the ratio of laser irradiance at that particular time, to the energy flux of the energetic electrons.

We note that the energy flux of the energetic electrons, $q_{ee}$ is the proportional to the integral of $(1/2)mv^3 f_1$, which in terms of our variables (recall $w \sim V^4$) is

$$q_{ee} \propto \int_0^\infty dw\, w^{1/2} f_1(w, z) \tag{9}$$

In this paper we calculate only the relative flux, since our assumption is that at some position, the quarter critical density here, the flux of energetic electron is known. From this knowledge, we can determine the flux everywhere once we have solved for $f_1(w,z)$.

Let us briefly discuss a comparison of the URLLE approach to the calculation of energy flux; and more particularly fuel and ablator preheat, to that derived here. They use a Monte Carlo technique. While they undoubtedly calculated the correct electron orbits as specified by their Monte Carlo formulation, their illustration, Fig. (10A) [taken from (15)] is one of straight line orbits for the electrons, indicating that they at least visualize a considerable distance of these electron orbits as straight line orbits.

Nevertheless the reality can be quite different from the illustration. First of all, as apparent from Fig (7B), an energetic electron at the thermal energy (here taken as 40keV) has a mean free path due to electron electron energy loss collisions which would take it from the quarter critical surface (r=1200, z=6.3) to (r=750, z=5.3). However as we have just said in the last section, the electron can only travel about ¼ of this distance due to the very rapid increase of the dynamical friction as its energy decreases. But that is the least of the issues. There are also electron collisions with both ions and plasma electron which scatter the electron in angle. These give an angular scattering rate (Z+1) times greater than the slowing down rate. Hence a more accurate illustration of the energetic electron orbit is as shown in Fig (10B), redrawn from (15), where the orbit is sketched first for only electron electron energy loss collisions, and then for an actual orbit in the plastic target (Z=5.3) and the silicon target (Z=14).

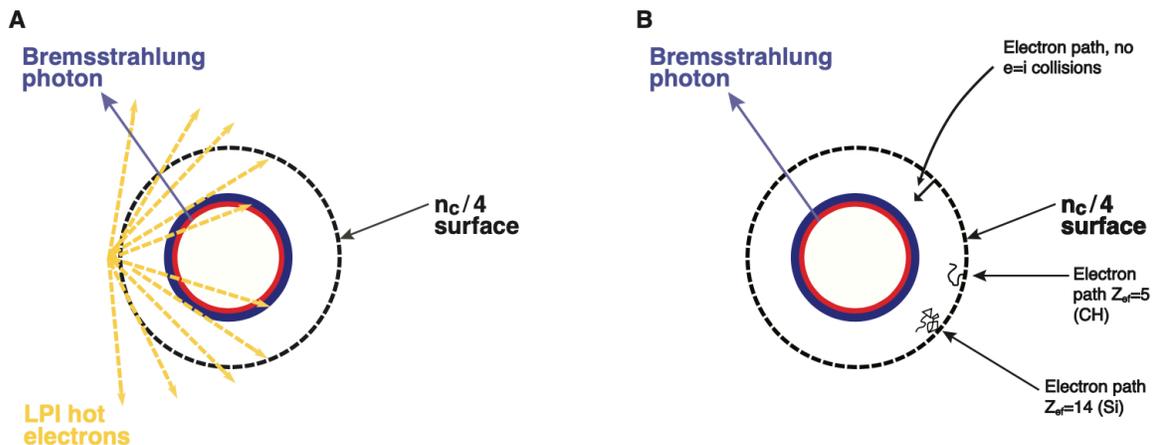

Figure (10). Sketches of electron orbits as envisioned for A: Monte Carlo trajectories, taken from (15), and B: for a Fokker Planck calculation.

Finally let us specify the distribution function we use as an initial value at the quarter critical density. Both the experiment and the simulation give a reasonable estimate of the hot electron temperature, which we take as 40 keV. But what about the angular distribution? We choose a distribution function

$$f(w,\theta,z_o) = \alpha u(\cos\theta)\exp{-u^2} \qquad (10)$$

where $u = V/T$ and of course, $w = u^4$. The coefficient $\alpha$ is determined by the energy flux at the quarter critical density. However this does not look very much like the distribution shown in Figs (5 and 6) and measured by URLLE just beyond the quarter critical surface. However Eq. (10) is a very reasonable choice, as discussed in Appendix B.

To continue, we discuss a point made in (4), namely that a sizable fraction of the energetic electrons produced may never make it into the target, because they reflex around the outer part of the blowoff plasma. This is discussed in more detail in Appendix C.

These experiments do show that an intermediate Z impurity in the target could have an effect of decreasing the energetic electrons. Another approach is the use of shorter wavelength lasers for instance excimer lasers, since the instability threshold generally goes as $I\lambda^2$. NRL has always used KrF lasers with a wavelength of 248 nm, and has recently begin to explore ArF lasers with a still shorter wavelength of 193nm (16).

V: Approximate Fokker Planck Solutions

We consider solving Equation (8) by separation of variables; let $f_1(z,w) = \Phi(z)\Omega(w)$. Then in the regions where Z is independent of z (or equivalently independent of r), Eq. (8) can be manipulated so there is only a function of z on the lhs, and only a function of w on the right. Hence each side must be a constant which we call $\kappa^2$.

$$\frac{d^2\Phi}{dz^2} + 2\frac{d}{dz}\left[\frac{\Phi}{k(z)r(z)}\right] = \kappa^2\Phi$$

(11)

$$\frac{d^2\Phi}{dz^2} + 2(r(z)k(z))^{-1}\frac{d\Phi}{dz} + 2\frac{d(rk)^{-1}}{dz}\Phi$$

$$= \kappa^2\Phi \quad (12)$$

and

$$\frac{d^2\Omega}{dw^2} - \frac{1}{4}\frac{1+Z}{w}\frac{d\Omega}{dw} + \frac{1+Z}{4w^2}\Omega = \kappa^2\Omega$$

(13)

Note that $\kappa^2$ is positive for the configuration we are considering and recall that k is a function of both r and $T_e$. As a function of w, $\Omega$ must approach 0 for large w; and as a function of z, $\Phi$ must increase from $\Phi = 0$ at z=0.

A. The solution for the spatial (i.e. z) equation:

First notice that in contrast to the diffusion approximation, the Z only appears in the energy equation, i.e. the $\Omega$ equation, and makes no appearance in the spatial equation.

We begin by considering the planar case. In a later section we discuss the effect of the spherical geometry. Neglecting the spherical term, i.e. the second term on the left of Eq. (12), the solution becomes particularly simple

$$\Phi \propto \sinh \kappa z$$

B. The solution for energy (i.e. $\Omega$) equation

This energy equation, Eq. (13), looks like an equation for a modified Bessel function, since $\kappa^2$ is positive. We then make a simple transformation of the independent variable to $\omega = \kappa w$, and a transformation of the dependent variable to $\Omega = \omega^n \Psi(\omega)$. We find that the equation for $\Psi$ is

$$\frac{d^2\Psi}{d\omega^2} + \frac{1}{\omega}\frac{d\Psi}{d\omega} - \left(1 + \frac{\alpha^2}{\omega^2}\right)\Psi = 0 \qquad (14)$$

For $n = (Z+5)/8$ and $\quad \alpha^2 = [(Z-3)/8]^2 \qquad (15)$

Equation (14) is the equation for the modified Bessel functions, so $\Psi(\omega) = I_\alpha(\omega)$ or $K_\alpha(\omega)$ or some combination of the two. However, the only one that satisfies the boundary condition as w goes to infinity is the K function. Hence the Z of the plasma is reflected only in the index of the modified Bessel function, and the power of $\omega$ which multiplies it.

Hence the solution is

$$\Psi(\kappa w) \propto K_{\left|\frac{Z-3}{8}\right|}(\kappa w) \qquad \Omega(w) = (\kappa w)^{\frac{Z+5}{8}} \Psi(\kappa w)$$
and $\qquad (16)$

Hence the solution for $f_1$ characterized by $\kappa$ is

$$f_1(w,z) \propto (\kappa w)^{\frac{Z+5}{8}} K_{\left|\frac{Z-3}{8}\right|}(\kappa w) \sinh \kappa z \qquad (17)$$

C: An expansion in 'sparse' eigenfunctions

Ordinarily, if one knows, let's say $f_1(w,z_o)$ where $z_o$ is the position of the quarter critical surface, one writes the $f_1(w)$ at that position as a summation of the different values of $\kappa$ that make up the function. Then one multiples each by the appropriate spatial dependence and for that K, and has the solution for all w and $z < z_o$. The problem is that this technique only works if the eigenfunction are orthogonal and complete.

The eigenfunctions expressed in Eq. (17) are not complete, and they are certainly not orthogonal to one another. They can hardly be orthogonal if each one is everywhere positive in both the spatial and energy domain. Hence one cannot construct a solution to Eq. (8) by constructing a unique expression for $f(w,z_o)$ as a sum or integral over the eigenvalues in the w domain as written in Eq. (17).

But what if $f(w,z_o)$ just happens to equal a constant times

$$(\kappa w)^{\frac{Z+5}{8}} K_{\left|\frac{Z-3}{8}\right|}(\kappa w)$$

?

Then one knows the solution for all w and $z<z_o$. Of course $f(w,z_o)$ rarely is so cooperative. However, what if it approximately equals a summation of a small number of such terms. As we have asserted, the solution in the entire range is stable to small perturbations on the boundary. Hence if we can solve it using an approximate expression for the boundary value, it should be a reasonably good approximation everywhere in the domain.

We find that this is a reasonable approximation. Hence, we denote this as a sparse eigenfunction expansion, since the number of eigenfunctions is small, and the match is at best, approximate. The distribution function this summation attempts to reproduce is of course $w^{1/4}\exp-w^{1/2}$. The integral for the energy flux can be written in the w (or $\zeta$, the energy divided by $T_o$) domain as

$$q \propto \int_0^\infty d\zeta\, \zeta^{5/2} \exp-\zeta \equiv \int_0^\infty dw\, w^{3/4} \exp-\sqrt{w}$$

(18)

Clearly, expressed in the $\zeta$ domain, the integrand has its maximum at $\zeta=5/2$, or w = 6.25. Thus, we want any sparse eigenfunction expansion to match $f_1(w,z_o)$ as well as possible around this value of w.

The algorithm we choose to match $f_1(w,z_o)$ is to first find an eigenfunction (i.e. a value of $\kappa$ and amplitude coefficient) that matches exactly at w = 3 and 6. Then take another eigenfunction that matches exactly at w = 6 and 10 and take half the sum of these two eigenfunctions. Of course, this will match exactly at 6. Then take the $f_1(w,z_o)$ minus the first two eigenfunctions as a function of w. For a third eigenfunction, take one that matches this construction at w = ½ and 3/2. This approach could very well be improved by applying machine learning.

For simplicity, we will consider only Z's that make the calculation as simple as possible. Namely we take Z = 3 as a stand in for plastic with a Z = 5.3, and Z = 11 as a stand in for silicon with a Z=14. Of course, we will always need to consider Z=1. The sparse eigenfunctions then are:

For Z = 1, $(\kappa w)^{3/4} K_{1/4}(\kappa w)$                          (19a)

For Z = 3: $(\kappa w) K_0(\kappa w)$                                  (19b)

For Z = 11: $(\kappa w)^2 K_1(\kappa w)$                             (19c)

The three term sparse eigenfunction expansion for Z = 1, is not necessary at this point, because for the region where z is so small the eigenfunction for the energy flux is simply linearly proportional to z, and all we need do is match it at the deuterium plastic boundary to get the energy flux into the fuel.

The sparse eigenfunction expansion for $f_1(w,z_o)$ for our substitute for the plastic (Z=3) is

0.23(.25w) $K_0$(.25w) + 0.27(.32w)$K_0$(.32w) + 57(.9w)$K_0$(.57w)

Actually, we find the reducing the first two terms gives a slightly better fit to the assumed distribution function $w^{1/4} \exp\text{-}w^{1/2}$. Hence we use

$$0.19(.25w) K_0(.25w) + 0.22(.32w)K_0(.32w) + .57(.9w)K_0(.57w) \qquad (20)$$

For the Silicon (taking Z=11) the sparse eigenfunction is

$$.16(.4w)^2 K_1(.4w) + 0.2(.48w)^2 K_1(.48w) + . + (1.8w)^2 K_1(1.8w) \qquad (21)$$

Plots of $f_1(w,z_o)$ and the sparse eigenfunction approximation of each are shown in Figs. (11 and 12).

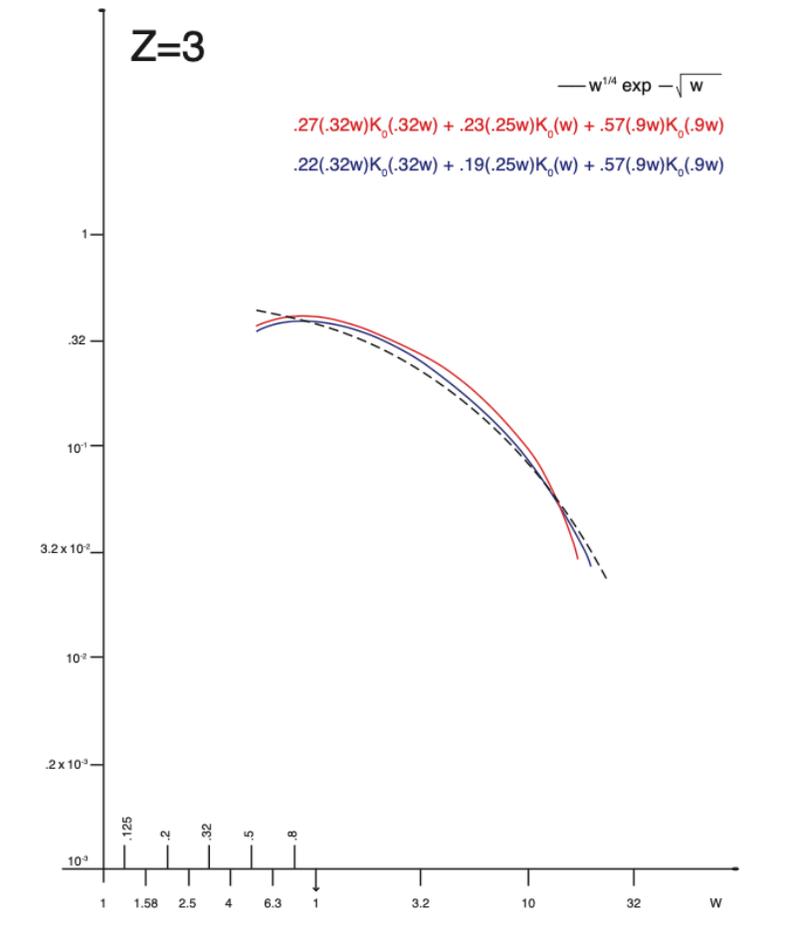

Figure (11): For Z=3, a plot of the distribution function, $w^{1/4}\exp{-w^{1/2}}$ in black dashed, the initial three term sparse eigenfunction approximation to it in red, and a slightly corrected version in solid blue. On the horizontal axis, the numbers go from 0.1 to 1, so the lower numbers should be divided by 10.

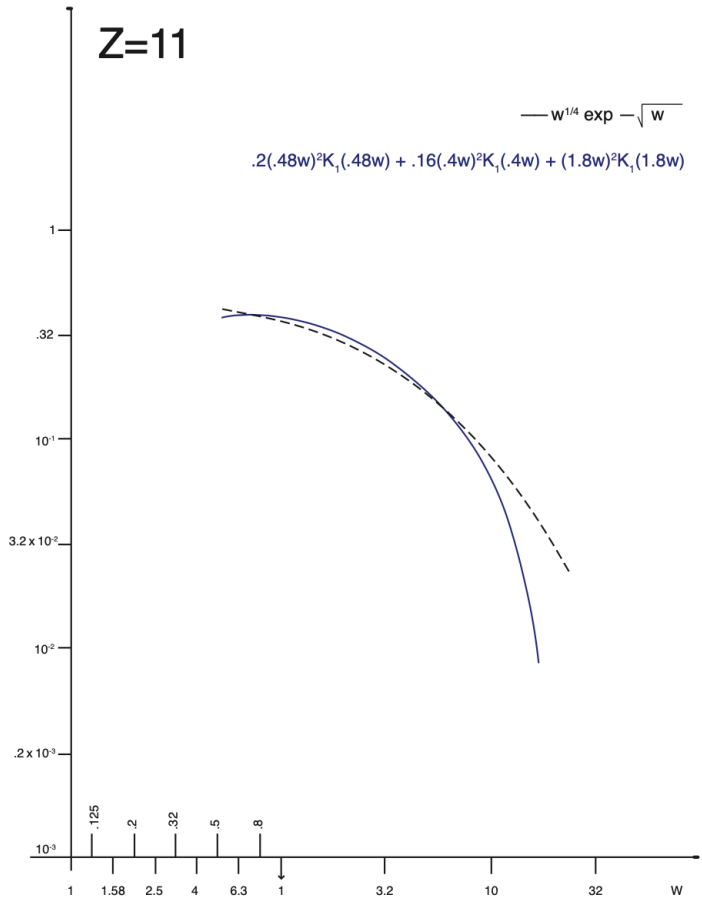

Figure (12): A plot of the distribution function, $w^{1/4}\exp{-w^{1/2}}$ in black dashed, and the three term sparse eigenfunction approximation to it in blue solid for Z=11. Between 0.1 and 1, the numbers should be corrected as in Figure 11.

There are several ways to check the validity of the sparse eigenfunction approach. First one might find another example, for which a known exact solution exists, and compare it with the sparse eigenfunction approach. This is done in Appendix A. Secondly one could make as accurate as possible a comparison with experiment. That is the focus of this paper. Finally, one could compare the sparse eigenfunction approach with a numerical solution of Eq. (3).

That would require a capability for Fokker Planck equation simulation. Neither the author, nor even the laser fusion group at NRL have such a capability at this point in time.

VI. Energetic electron energy flux in a segmented target.

As we have seen, in a laser target experiment, the electron electron collision time is very, very short compared to the implosion time. Hence for the course of the implosion, the energetic electron distribution function can be regarded as being in a steady state at each instant of the implosion. The quantity of interest is then their instantaneous energy flux and the energetic electron temperature. Since these energetic electrons only lose energy by colliding with the plasma electrons, the divergence of their energy flux gives the preheat of the plasma electrons by these energetic electrons.

The experiments cited give a different measurement. These are experiments at basically a single irradiance for a time of several nanoseconds (see Fig. (2)). From the hard X-ray measurement, it comes up with a total energy and temperature of the fast electrons over the entire pulse. It then compares this total energy to the total laser energy in the pulse. By doing this for different laser energies, one can develop an empirical law for the total energetic electron energy and temperature as a function of total laser energy. Our assumption is that the plasma reaches an equilibrium very quickly on the time scale of the laser pulse, so that the ratio of energy flux of energetic electrons to instantaneous laser irradiance is the same.

Let us imagine that the target is composed of various components, each bounded at a particular radius. We are interested only in radii (or z) less than the quarter critical radius, for the particular laser plasma instability we focus on. For instance the target with the electron density given by Fig (7A) is composed of 3 regions, Si from 1200 μm (the quarter critical radius) to 900; CH from 900 to 600, and D from 0 to 600. Let us use the subscript i to denote the region of the target and j to denote the particular sparse eigenfunction. Let us count the regions of the target starting with 1 for the region adjoining the quarter critical surface, and counting upward until we reach the fuel. Hence for the silicon target (see Fig (7A)), the silicon is region 1, the plastic, 2 and the deuterium, 3. Each sparse eigenfunction is characterized by two parameters, its coefficient a, and it wave number κ. We will count the highest κ as sub 1 (corresponding to the sparse eigenfuntion centered around the lowest energetic electron energy) and the lower κ's counting upward. For instance the $κ_{2,2}$ corresponds to the plastic portion of the silicon target; and second, the middle value of κ, or a numerical value of $\kappa = 0.48$. The $a_{1,2}$ is the coefficient of the middle term of he silicon target, or $a_{1.2} = 0.2$ (see Eq. (20)).

Hence let us take the distribution function in the i'th region of the target

$$f_i(w,z) = \sum_j a_{i,j} \Omega_{i,j}(\kappa_{i,j} w) \Phi(\kappa_{i,j} z)$$

(22)

Where the Ω's are given in Eq. (16) and for now we assume a planar approximation which vanishes at z (i.e. r) =0. Starting with the portion of the target adjoining the quarter critical density at $z_o$,

$\Phi(\kappa_{1,j} z) = [\sinh \kappa_{1j} z] / [\sinh \kappa_{1j} z_o]$  (23)

Recall that at $z=z_o$ the distribution function was taken simply as $f = w^{1/4} \exp{-w^{1/2}}$. That is at this point, we are not concerned with how energy is normalized, but are concerned with only the functional form of the distribution function. Actually we will calculate only the ratio of the energy flux at z to that at $z_o$, so we do not have to provide any normalization for that part of the calculation.

The energy flux is given by Eq. (9), or for the sparse eigenfunction expansion

$$q_i(z) \propto \int_0^\infty dw\, w^{1/2} \sum_j a_{i,j} \Omega_{i,j}(\kappa_{i,j} w) \Phi(\kappa_{i,j} z) \equiv \sum_j a_{i,j} \kappa_{i,j}^{-3/2} \Phi(\kappa_{i,j} z) \left[ \int_0^\infty du\, u\, \Omega_{i,j}(u) \right]$$

(24)

Note that whatever the sparse eigenfunction Ω(w) is, the integral over w in calculating the energy flux is the same for all κ, once the integral over w is rewritten in dimensionless form. Hence in comparing the ratio of the energy flux at a particular z to another z, the integral over the energy part of the distribution function cancels out.

Then for the region adjacent to the quarter critical surface, we can easily form the ratio

$$\frac{q(z)}{q(z_o)} = \frac{\sum_j a_{1,j} \kappa_{1.j}^{-3/2} \frac{\sinh \kappa_{1,j} z}{\sinh \kappa_{1,j} z_o}}{\sum_j a_{1,j} \kappa_{1.j}^{-3/2}}$$

(25)

To get the actual value of the energetic electron flux in region 1, adjacent to the quarter critical density, simply multiply the ratio above, by the energetic electron energy flux at the quarter critical density.

The next issue is to transfer from one region to another, let's say the silicon to the plastic, at z defined as $z_2 < z < z_1$, where $z_1$ is the boundary between the silicon and plastic, and $z_2$ is the boundary between the plastic and deuterium. Clearly there is neither a source nor sink of energetic particles at $z_1$, and there is no reason to believe any energetic particles are reflected from this interface, so the nonlocal electron energy flux must be continuous across this boundary and equal to the incident energy flux.

Hence from Eq. (25) and the given source at $z=z_o$, first find the energy flux at $z=z_1$. Then work backwards toward $z=0$ with this as the the starting point, analogous to the known flux at $z=z_o$.

For the last region; the fuel, the outer boundary of the fuel, $z_2$, is so small that all of the sparse eigenfunctions are in the linear regime $\sinh \kappa z = \kappa z$. Hence whatever the energy flux is at $z_2$ (or whatever the boundary of the fuel is if there are more than 2 layers between the fuel and quarter critical density), inside the fuel the energy flux as a function of z, just drops to zero linearly at $z=0$.

We briefly discuss the case of the spherical configuration and the effect of the higher order terms in the Legendre expansion in later sections.

## VII. Legendre Expansion

In our previous development, we assumed a 2 term Legendre expansion, but never questioned it. Here we will look at it a bit more carefully and look at the conditions for neglecting higher order terms. Here we examine that assumption, by seeing approximately the correction to the $f_1$ term (i.e. the term responsible for the energy flux) generated by including the $f_2$ term.

First a brief review of some of the features of Legendre polynomials. They form a complete orthogonal set, so that

$$\int_{-1}^{1} d\mu P_n(\mu) P_m(\mu) = \frac{2}{n+1} \delta_{mn} \tag{26}$$

Also there are several recursion relations which are useful

$$(n+1) P_{n+1}(\mu) = (2n+1) P_n(\mu) - n P_{n-1}(\mu) \tag{27}$$

$$\frac{d}{d\mu} P_{n+1}(\mu) = (n+1) P_n(\mu) + \mu \frac{d}{d\mu} P_n(\mu) \tag{28}$$

$$\frac{d}{d\mu}(1-\mu^2) \frac{dP_n}{d\mu} = -n(n+1) P_n \tag{29}$$

The first four Legendre polynomials are

$$P_o = 1 \qquad P_1 = \mu \qquad P_2 = \frac{3}{2}\mu^2 - \frac{1}{2} \qquad P_3 = \frac{15}{6}\mu^3 - \frac{9}{6}\mu$$

Then expressing the steady state Fokker Planck equation as a summation over Legendre polynomials, we find it is

$$\sum_n \left\{ V\mu \frac{\partial f_n}{\partial r} P_n(\mu) - \frac{V}{r}(1-\mu^2) f_n \frac{dP_n}{d\mu} - \frac{1}{V^2} \frac{\partial}{\partial V} V^3 v f_n P_n + n(n+1)(1+Z) v f_n P_n \right\} = 0$$

(30)

The next few steps, which are not worth detailing here are:

1. Make the transformation of variables from (r,V) to (z,w)
2. Use the recursion relations for the Legendre polynomials so that they are all just simple $P_n$'s (i.e. no derivatives, no multiplication by various powers of μ)
3. Note that all f's above have the same index. Convert the summation to one where all P's have now the same index, but the elements of the summation couple different values of f with different n's.
4. Multiply the equations first by $P_o$, $P_1$, and $P_2$; integrate over μ from -1 to 1, and neglect all f's of order 3 and higher.
5. Neglect the spherical terms (those with an $r^{-1}$, these effects will be treated in the next section).
6. Assume the terms arising from $f_2$ are small and use perturbation theory.

The equations for $f_o$, $f_1$ and $f_2$ become:

$$\frac{1}{\sqrt{3}} \frac{\partial f_1}{\partial z} = \frac{\partial f_o}{\partial w}$$

(31)

$$\sqrt{3} \frac{\partial f_o}{\partial z} + 2 \frac{\sqrt{3}}{5} \frac{\partial f_2}{\partial z} = \frac{\partial f_1}{\partial w} - \frac{1+Z}{4w} f_1$$

(32)

$$\sqrt{3}\frac{2}{3}\frac{\partial f_1}{\partial z} = \frac{\partial f_2}{\partial w} - \frac{3}{2}\frac{1+Z}{w}f_2$$

(33)

Note that the last of these 3 equations is a simple first order linear inhomogeneous equation for $f_2$ if $f_1$ is known. Then the solution for $f_2$, which vanishes as w approaches infinity is

$$f_2 = -w^B \int_w^\infty dw'(w')^{-B} \frac{2\sqrt{3}}{3}\frac{\partial f_1(w',z)}{\partial z}$$

(34)

where B=3(1+Z)/2. For Z=3, B=6, and for Z=11, B=18. So in either case, the integrand is dominated in the region pf w space around w=w'. Making this approximation we can approximate the integral as

$$f_2 = -\frac{w}{B-1}\frac{2\sqrt{3}}{3}\frac{\partial f_1}{\partial z}$$

(35)

and insert this into the second equation to get an equation for $f_o$ and $f_1$

$$\sqrt{3}\frac{\partial f_o}{\partial z} + \frac{2\sqrt{3}}{5}\frac{\partial}{\partial z}\left[-\frac{w}{B-1}\frac{2\sqrt{3}}{3}\frac{\partial f_1}{\partial z}\right] = \frac{\partial f_1}{\partial w} - \frac{1+Z}{w}f_1$$

(36)

We want to solve Eq. (36) within the framework of the sparse eigenfunction method. In Eq (36), the second term on the left hand side, that which was created from the $f_2$ term is treated as a small perturbation, so that one can replace the second partial derivative with respect to z with $\kappa^2$.

Using the first equation, then one can get an equation for only $f_1$. It becomes

$$\frac{\partial^2 f_1}{\partial z^2} = \frac{\partial^2 f_1}{\partial w^2} - \frac{1}{4}\frac{\partial}{\partial w}\left(\frac{1+Z}{w}\right)\left[1 - \frac{8\kappa^2 w^2}{5(Z+1)(3Z+1)}\right]f_1$$
(37)

Note that Eq(37) with the correction from $f_2$ has the same form as Eq (8), namely $f_1$ can be separated into a product of a function of only z times a function of only w, i.e. $f_1(w,z) = \Phi(z)\Omega(w)$ as previously. In fact, if necessary the spherical correction can be added to the left hand side, i.e. the z dependent part. Then one finds two equations for the single variables $\Phi$ and $\Omega$. The equation for $\Phi$ is unchanged, and still has the same solution in the planar approximation

$\Phi(z) = \sinh \kappa z$

Hence any change in the z dependence arises only from changes in the κ's generated by the $f_2$ effect. However the equation for $\Omega$, for each κ, is no longer an equation for modified Bessel functions of the second kind, but is some other function. One choice could be to solve for these function and reconstruct $f(w,z_o)$ as a summation over these new functions of w. Alternatively one could numerically solve Eq.(37) as a second order partial derivative equation.

However we will take a simpler, more approximate approach based on the way in which $f_1(w,z_o)$ is constructed over a summation of sparse eigenfunctions. Note that the the sparse eigenfunction goes to zero as w approaches 0 and infinity, and does so much faster than $f_1(w)$ approaches these limits. Hence there is only a limited value of w where the sparse eigenfunction is a good approximation for $f_1(w)$. For values of w below and above this range, the sparse eigenfunction is much smaller than $f_1$, and as long as it remains so, any poor approximation in these outer regions will not affect the final result. As one moves from lower to higher w, one moves from one sparse eigenfunction with larger κ to another with smaller κ. This is illustrated in Fig.(13)

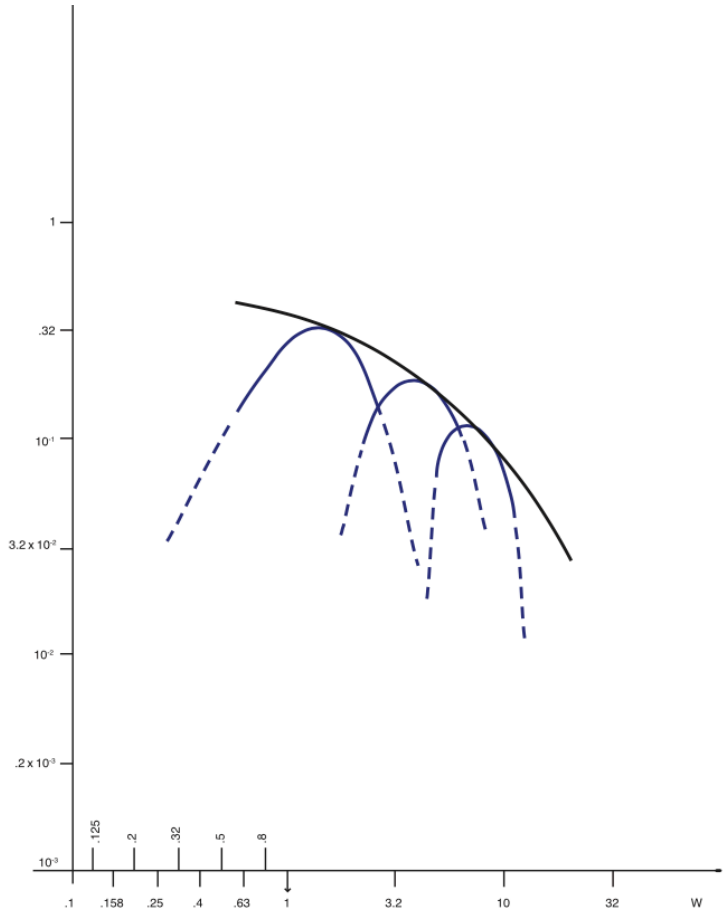

Figure (13) The thicker curve is $f_1(w,z_o)$. The three other curves are schematics of sparse eigenfunctions centered at w~1.5, w~3-6, and w~6-10, as in our case for Z=3. The darker region of these curves are where the sparse eigenfunctions are reasonable approximations for $f_1$, the dashed regions are where they are much smaller than $f_1$.

Hence in Eq. (37), we approximate the value of κw to the region where it is a good approximation for that particular sparse eigenfunction. To be specific, we consider the case of Z=3, and the particular sparse eigenfunction centered between w = 3 and 6. For that, case the unperturbed κ is 0.3. Using Eq. (37), let us find the correction to κ (=κ+δκ) driven by $f_2$. We take κw = 1.5. Then Eq. (37) is still an equation for a power of κw times a modified Bessel function of the second kind. However the power of κw is shifted, as is the order of the Bessel function.

The Bessel function solutions are now $(\kappa w)^m K_{|\alpha|}(\kappa w)$ where

$$|\alpha| = \left| \frac{Z-3}{8} - \epsilon \frac{Z+1}{8} \right|$$

(38)

$$m = \frac{Z+5}{8} - \frac{Z+1}{8}\epsilon$$

(39)

and

$$\epsilon = \frac{18}{5(Z+1)(3Z+1)}$$

(40)

Notice that for Z=1, the perturbation expansion is likely qualitatively correct, but not a very good approximation, i.e. $\epsilon$=0.45. However in a real laser implosion, the region where Z=1 is very small (as a function of z), whereas the outer regions are characterized by higher Z (i.e. plastic Z=5, or silicon, Z=14)in the NIF/URLLE spherical cases we are considering). For Z=2, $\epsilon$=0.17, so the expansion ought to be reasonably good, but there will probably be significant effects from the extra term in the Legendre expansion.

Let us consider the case for Z=3, where $\epsilon$ =0.09, and the sparse eigenfunction has an unperturbed $\kappa$= 0.3. The corrected sparse eigenfunction now becomes $(\kappa w)^{0.955} K_{0.045}(\kappa w)$.

The main effect of the $f_2$ is in the power of $\kappa w$ multiplying the Bessel function. For the values of $\kappa w$ that we are interested in, the difference between $K_{0.045}(\kappa w)$ and $K_o(\kappa w)$ is typically beyond the third decimal place, for the arguments of concern here, according to the Keisan on line calculator (https://keisan.casio.com/exec/system/1180573476). Hence we consider only the effect of $(\kappa w)^{0.955}$. Doing a calculation like that in Section V, we find that $\delta\kappa = -0.02$, or about 7% of the unperturbed $\kappa$ for the particular case we consider. Notice that the sign is negative, meaning that the damping is a bit less than the two term Legendre expansion would give. In other words, including the $f_2$ gives heating by energetic electrons at further distances from where they are produced, than does the simple two term expansion.

Hence for Z=3, we consider that the 2 term Legendre expansion of the Fokker Planck equation, may give rise to an over estimate of the damping of ~7%. At this point, we adopt this estimate as a rule of thumb. However for the case of Z = 11, $\varepsilon$=0.009, so that in this case, the two term Legendre expansion should be very good.

VIII Spherical effects

Using our sparse eigenfunction formulation, the equation for $f_l(w,z)$ is given in Eqs.(8). Notice that the charge state Z, once again, appears only in the equation for $\Omega$. However the spherical effects appear only in the equation for $\Phi$, the equation for the spacial dependence. Hence once k(z) and r(z) are known (neither of which depends on Z), the dependence of the spatial profile on Z comes in only through the particular values of $\kappa$, which are determined by the Z. This is also true of the corrections from higher order terms in the Legendre expansion. Hence the solution for $\Phi(z)$ is basically the same as in our earlier work where we considered only Z=1 (3), as long as the proper values of $\kappa$ are chosen for the appropriate Z's. The form of $(kr)^{-1}$ is rather complicated, and it is difficult to get analytic results, as pointed out in (3). However the numerical calculation shows several effects. First of all, it shows that the planar model is reasonably accurate, even for the case of instability generated energetic electrons. The main difference is the presence of an additional barrier for the electrons around the fuel due to the very large density gradients there. This extra barrier reduces the fuel preheat by about a factor of 2 as Fig (14), taken from (3) shows. It is reproduced here as Fig (14). However these differences are all basically independent of Z, except for the dependence of $\kappa$ on Z. However in the fuel region, where z<<1, as we will see, $\Phi(z) \sim z$ with virtually no dependence on $\kappa$. Furthermore the numerical calculation shows that there is less damping in the outer region, perhaps due to some compression of the energy flux as it moves inward. This also gives rise to somewhat greater flux (perhaps ~ 10%) in the outer regions of the ablating plasma, but once near the fuel, the flux decreases from that of the planar model.

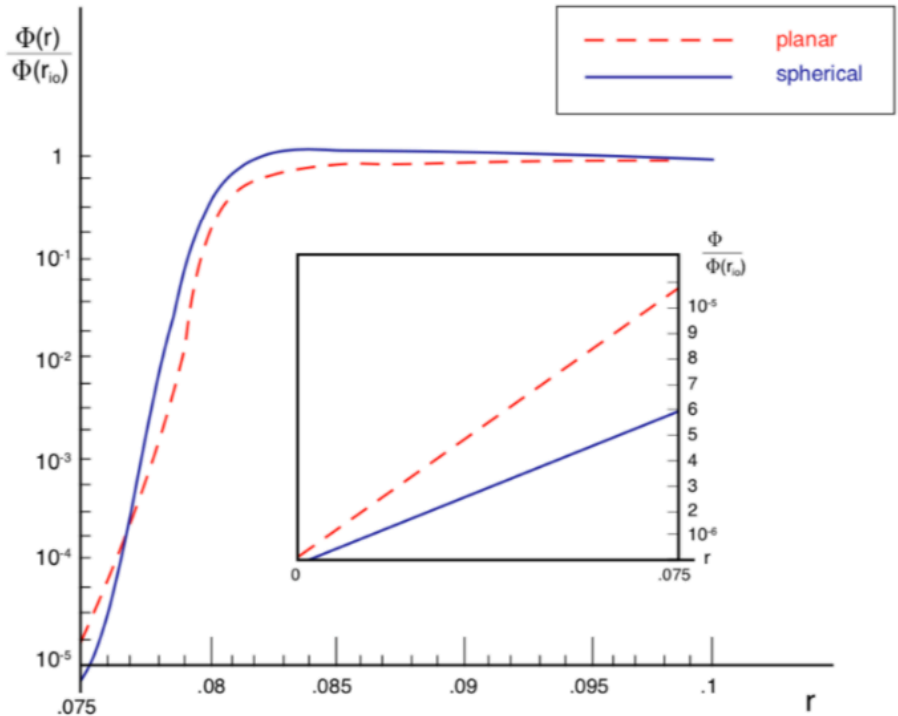

Figure (14); This shows the numerical calculation for flux for instability generated energetic electrons, comparing the planar and spherical calculation, taken from (3).

Figure (15) shows graphs of $(kr)^{-1}$, $z(r)$ for for the density profile for the URLLE/NIF experiment, as shown in Fig. (7A), as well as a typical value of κ for the profiles used here. Notice that the region r<0.6, and r>0.8 are dominated by spherical effects, while the region between, where the damping is maximum, is dominated by the large dissipation in the dense plasma region. However as Fig. (14) shows, this does not make a significant difference in total energy flux of the instability generated electrons, and as pointed out in (3), the actual difference can only be determined accurately by a numerical integration. Hence we will confine our remarks to a few qualitative issues.

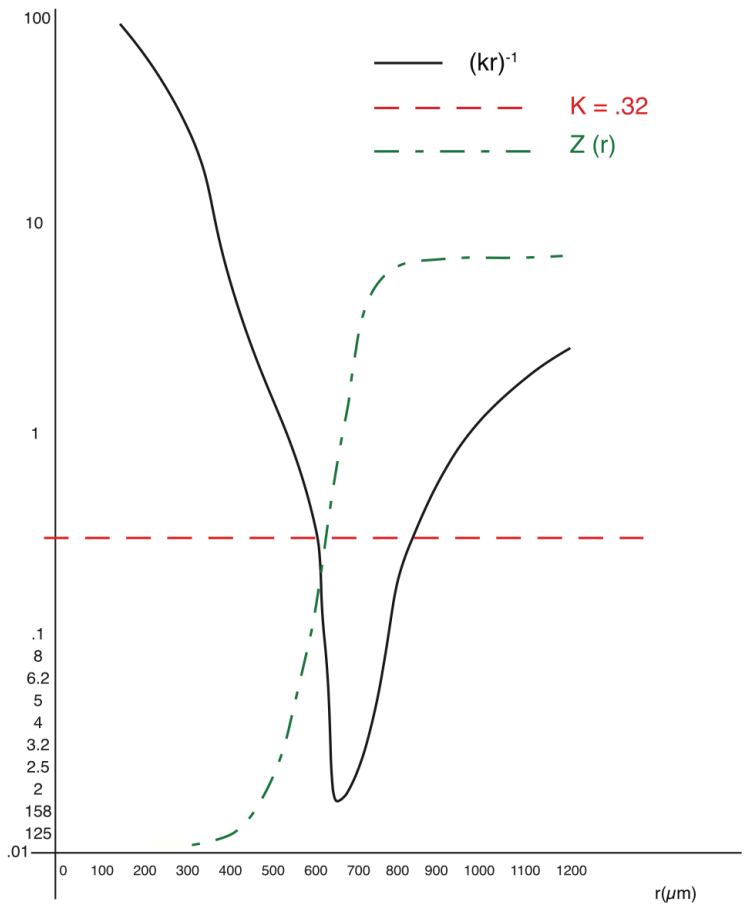

Figure (15): Plot of $(kr)^{-1}$, solid black; z(r) green dash dot; and $\kappa = .32$, red dash a typical value for the URLLE/NIF results for Z=3.

In the planar approximation, Eq. (12) in the region of the origin as the two linear independent solutions

$\Phi \sim z$   and   $\Phi = $ constant

Since $\Phi$ must approach zero as z ( and of course r) approaches zero, only the first solution is physically reasonable and the second solution is unphysical and must be eliminated.

Now consider Eq. (12) with the spherical term included.  For very small z and r, the density is constant, approximately for the URLLE/NIF experiment, exactly for the NRL calculations.  Hence we take kr=z in this region near z=0.  The equation has the form where a polynomial expansion is valid, certainly for small z.  We find that the two linearly independent solutions are

$\Phi \sim z$ and $\Phi \sim z^{-2}$.

Only the first solution is physically reasonable for small z, and the second must be eliminated.   Hence the planar and spherical solutions near z=0 have different unphysical solutions which must be eliminated.  However they have the same physically permitted solution.  Since in our cases, the fuel extends only to very small z's, both the planar and spherical solution will give a flux which is linear in z up to whatever it is at the inner edge of the the dense plasma bounding the fuel.

IX  Application to the URLLE/NIF experiment

We will now apply this theoretical development to the URLLE/NIF experiment. However we emphasize that we are only performing a qualitative, conceptional comparison.  First let us look at the experimental results on deposition profiles on targets with and without the silicon layer.  In Ref. (4), the authors present experimental measurements of the deposition profile, which they call the spatial dependence of the fraction of energetic electrons.   They arrive at this data by a combination of several (usually 2 or 3) shots on mass equivalent targets with germanium layers at various thicknesses and positions, deconvolving that data, and also by Monte Carlo simulations of a measured hot electron distribution at the quarter critical surface.   Figure (16) is taken from Ref. (4).

A

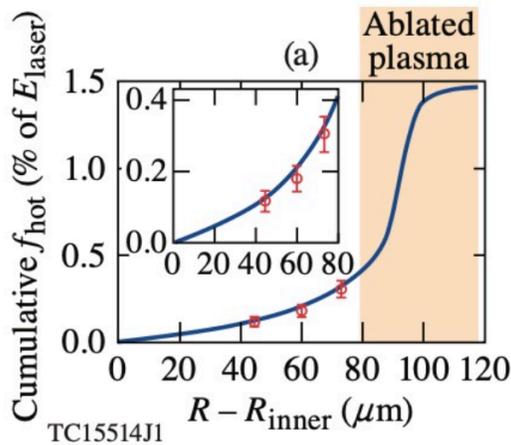

B

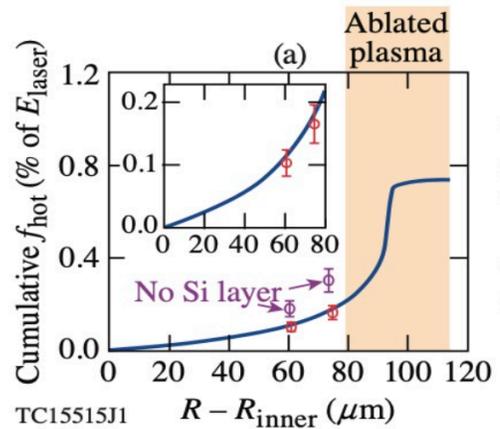

Figure (16):  A:  The measured spatial profile of the % of hot electron energy compared to laser energy for the pure plastic plasma.  The inset is a larger scale graph of the portion of the main graph from 0 to 80.  B: A comparison of the deposition with the silicon layer, showing that the silicon layer reduces the energy electrons by about half.  As shown in Figure (7), once the silicon

ablates, it becomes the entire plasma from the quarter critical surface to about 900 microns.

We find that the difference between the results of energetic electron production with and without the silicon layer is mainly due to the difference in density gradient at the quarter critical surface, as URLLE asserts. This difference is hardly even visible in the density profiles provided, but careful examination of the LILAC simulations confirm this. There is also a difference in energetic electron deposition profile due to the higher collisionality of the silicon plasma as compared to the CH plasma, but this difference, we find is less than the difference in energetic electron production at the quarter critical density due to the larger density gradient of the Si plasma. The URLLE/NIF experiment found that the total energy of the energetic electrons in the plasma with the silicon layer was 0.2% of the laser energy, and was 0.4% of the laser energy for the CH plasma. As mentioned we interpret this to mean that the ratio of energy fluxes as well as energy.

Getting the deposition profile (i.e the energy flux as a function of z) is then simply a matter of using Eq. (25) in the various regions. Since the author has no ability in this work to make a detailed comparison with (4), we simplify the calculation as much as possible, mainly to show that the theory gives a reasonable result. First of all we assume an energetic electron temperature of 40 keV, a bit less than the measured value of 50-55 keV. Hence we assume the plasma is a bit more collisional than it actually is. Secondly, to partially compensate for this error, we assume the carbon plasma has Z=3, instead of the actual value 5.3, and the silicon plasma has a Z=11, instead of the actual value of 14. This allows us to use simple modified Bessel functions, rather than functions with unusual indices, and we take $\Lambda = 8$ as indicated by Fig 9.. To reemphasize, the goal here is not to precisely, or even approximately fit the experiment, but to demonstrate that the theory, at this point, seems reasonable. Over the past year and a half, Andrey Solodov, Michael Rosenberg and I exchanged well over 40 emails, which the author appreciates very much. They certainly did as much as they could to help. However even with all this contact (but no lab visits, likely due in part to the pandemic), the author does not feel confident in claiming any more than that the approach, at this point, seems like a reasonable one.

We start with the known energy flux at the quarter critical surface and find the spatial dependence of the flux throughout that region. At the boundary of the next inner region, we set the energy flux equal to what it is at that boundary and work inward, continuing through all the regions, until we come to the fuel, at which point, we know that the energy flux simply decreases linearly in z to z=0 (i.e. r=0) as both the planar and spherical solutions would specify.

To start, let us once again list the a's and $\kappa$'s for both the plastic and silicon:

Plastic:  $a_1$=0.22,  $a_2$= 0.19,  $a_3$ = 0.57;  $\kappa_1$ = 0.32,  $\kappa_2$ = 0.25,  $\kappa_3$ = 0.57

Silicon:  $a_1$=0.2,  $a_2$= 0.16,  $a_3$ = 0.7;  $\kappa_1$ = 0.48,  $\kappa_2$ = 0.4,  $\kappa_3$ = 0.1.8

Then a simple application of Eq. (25) gives that the flux between 1200μm>r>900μm decreases by 9% in the plastic, and 18% in the silicon. Hence the presence of the silicon does provide additional reduction of electron energy flux, but it is a much less significant effect than the effect of the steeper gradient and increased collisionality on the instability.

For 600< r  900, it is all plastic and the rate of decay of the flux is the same for both targets. However we have seen in Section VII that the presence of higher order terms in the Legendre expansion and corrections due to spherical effects, the latter only roughly estimated from earlier numerical solutions, may reduce the decay rate to somewhat lower than the $\kappa$  in the plastic, while there is hardly any reduction for the silicon from higher order terms in the Legendre expansion. Figure (17) shows the application of Eq. (25) to the URLLE/NIF experiment.

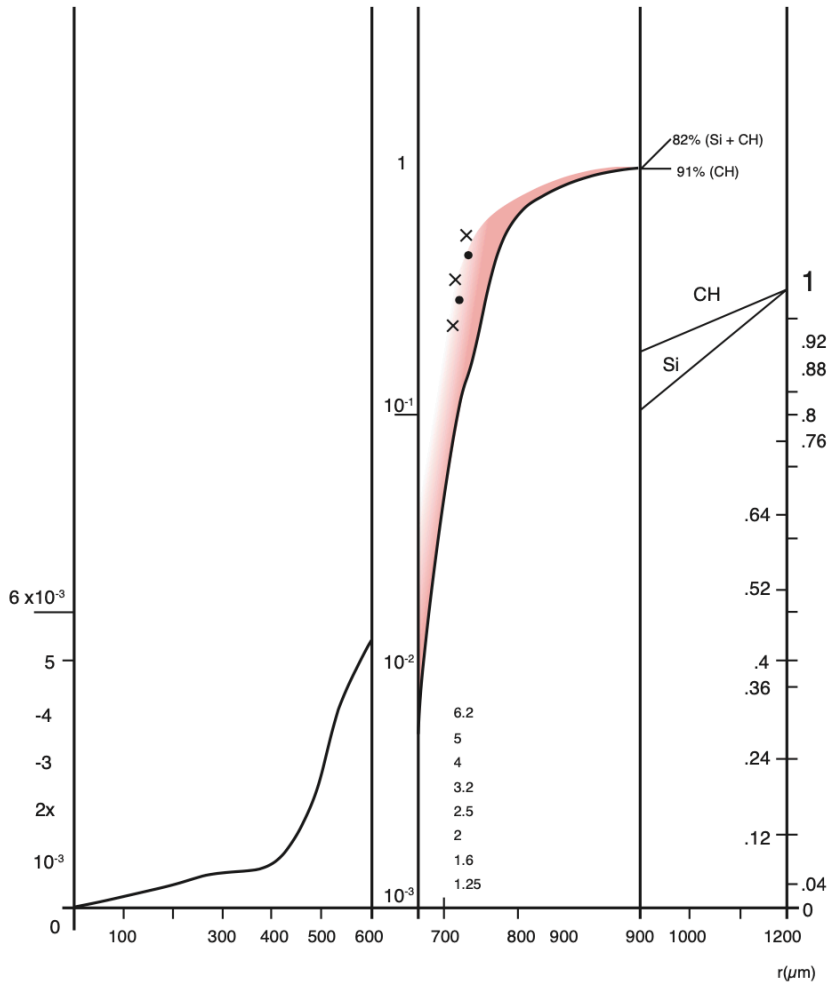

Figure (17)   A graph of the calculated energy flux of the instability generated electrons, which shows the result of applying Eq. (25) to the URLLE/NIF experiment, along with the data points for their plastic target and the plastic target with the silicon layer (4).  There are 3 regions in the graph.  From 900 microns to 1200, (the plasma blowoff) it is a linear scale from zero to one.  The right hand intercept (i.e. unity) is the assumed energy flux of the instability generated electrons, 0.4% of the laser flux at the quarter critical density for the plastic target, 0.2% for the target with the silicon layer.  This flux decreases at 900 microns to 91% for the CH target, and to 82 % for the silicon layer target.  From 600 microns to 900, (the ablating plasma) it is a log scale, where unity on that right hand intercept means the flux at 900 microns.  The two dots are the experimental results from the target with the silicon layer; the 3 crosses, the pure CH target. The pink region shows the result of reducing the dissipation by ~10%, as the inclusion of the extra term in the Legendre expansion and spherical effects suggestion it might be. Finally the region from 0 to 600 microns (the fuel) is plotted separately on a linear scale. As a function of z the decrease in flux would be linear, but in r it is distorted by the nonlinear relation between z and r.

The experiment did not discuss any measurement of the fuel preheat, at least not yet. We consider the experiment where the laser flux is $10^{15}$ W/cm$^2$, as shown in Figure (4). For the silicon layered target, the instability generated energetic electron flux is 0.2% of that, $2\times10^{12}$ W/cm$^2$. As Figure (17) shows, from 1200 to 900 μm, where it is a silicon plasma, the energetic energy electron flux decays to 82% of its value at the quarter critical density. Then in the plastic plasma between 600 and 900 μm, it decays by another factor of $5\times10^{-3}$, so the energetic electron energy flux incident on the fuel is $8\times10^{9}$ W/cm$^2$. Since the energy flux is zero at the origin, this entire incident energy flux is dissipated in the fuel, i.e. the region from 600 μm to zero. Assuming average particle energy is 1.5 times the temperature, and assuming the average density of the DT fuel is ~ $3.5\times10^{21}$ cm$^{-3}$ (see Figure 7A), we find that the fuel heating is ~ 0.44 eV/ns, according to the planar model. However as Figure (14) shows, the 'spherical barrier' reduces this by about a factor of 2, so this theory would predicts a heating rate of ~0.22 eV/ns for the silicon layered plasma.

For the pure plastic plasma, as Fig (4) shows, the energetic electron flux is $4\times10^{12}$ W/cm$^2$. In the now plastic plasma between 900 and 1200, the flux decays to ~ 91% of its value at quarter critical. Between 600 and 900, it decays the same $5\times10^{-3}$, since in this region it is a plastic plasma in either case. Hence for the pure plastic ablation target, the heating rate from the energetic electrons is ~ 1eV/ns in the planar calculation, and this is reduced by the spherical barrier to ~ 0.5 eV/ns.

The silicon layer reduces the fuel preheat by a factor of 2 due to the fact that it produces fewer energetic electrons. This is due to the effect on the instability in the blowoff plasma; the threshold is increased due to the increased density gradient at the quarter critical surface, as well as from the additional dissipation there due to its higher Z. This factor of 2 reduction is further enhanced by the additional dissipation in the silicon, as opposed to CH portion of the plasma between 1200 μm and 900.

Of course this is only a rough calculation. However I do believe that this method of evaluating energetic electron deposition could give valuable information, both on the ablation layer spreading and the fuel preheat, if incorporated into a rad hydro code like FAST or LILAC.

X. Conclusions

This work extended the steady state Fokker Planck theory developed in (1,2) (the planar approximation for Z=1 plasmas) and (3) (the spherical version), to plasmas of arbitrary Z. The theory aims to develop a methodology for calculating the nonlocal electron energy flux for energetic electrons generated by a laser plasma instability. In a future work, the author intends to extend the theory for the uncoupled tail of a high temperature Maxwellian distribution in a plasma with arbitrary Z. It would also be interesting to investigate whether a similar approach would be useful for nonthermal ions. However the Fokker Planck equation for the fast ions, as opposed to fast electrons, is sufficiently different that no simple comparison is possible at this time.

While at this point the theory developed would not be applicable for plasmas with an arbitrary dependence of Z on radius, like that for say a hohlraum plasma, it is applicable for a segmented target with 2 or 3 different materials with different Z's, like those targets used in the URLLE/NIF spherical experiments For a plasma like a hohlraum plasma, where Z is a strong function of z, a possible approach would be to solve Eq. (8) in the two dimensional appropriate region of (w,z) space. The theory made use of a an approximate technique, new as far as the author is aware, which he called sparse eigenfunction expansion. Results so far indicate that it could be a very effective tool for problems of this type. The goal of the theoretical work was to come up with a theory accurate enough to be useful, but also simple enough, that it could be used at every fluid time step of a radiation hydro code. The steady state Fokker Planck theory with a sparse eigenfunction approximation to solving it appears to fit this requirement.

It shows that the steady state Fokker Planck model is not only a much better approach than the many Krook models (including the author's), but it gives much less fuel preheat than the Krook model. This is for a very simple, easily explainable reason, namely that the dynamical friction (which is not included in Krook models) increases very rapidly as the electron energy decreases.

It argues that steady state Fokker Planck theory is a better approximation than the so called time dependent two fluid 'diffusion' approximation. While the latter provides a simple scaling for plasma length scales with Z, the former does not. However after a somewhat more complicated calculation involving modified Bessel functions of the second kind of various orders, it does show a dependence of length on Z which is qualitatively similar.

This paper was also motivated by the URLLE/NIF experiment on spherical targets illuminated by the NIF laser. While detailed comparisons are not possible at this point due to further refinements which might be necessary for the theory and the inclusion of it in a rad hydro code, at this point it appears to be a very good start. A comparison of the calculated spatial decay of the energetic electron, as compared to the URLLE measurements are shown in Figure 17. The theory used very approximate values for many parameters in the experiment, some of which (i.e. the Z) are in fact well known, others (i.e. the $\Lambda$) are not so well known. However the reasonable agreement between theory and experiment does give encouragement for the theory at this point.

Appendix A: The characteristic method, A quick way to get a very rough solution for $f_1(w,z)$

Here we briefly sketch out another way to estimate the preheat from energetic electrons, whose deposition cannot be calculated only locally. This method we call the characteristic method. It is almost certainly less accurate that sparse eigenfunction, at least for the problem at hand. However it is much simpler and easier to use, it could well be a way of quickly getting a first approximation.

First neglect the spherical term in Eq. (8). Then replace the dependent variable $f(w,z)$ with $g(w,z)$ defined by

$$f_1(w,z) = w^n g(w,z)$$

(41)

Pick n = Z+1 so as to eliminate the first derivative term on the right and get

$$\frac{\partial^2 g}{\partial z^2} = \frac{\partial^2 g}{\partial w^2} - \frac{(1+Z)(9+Z)}{64w^2}g$$
(42)

The characteristic method assumes that the second term on the right is negligible (i.e the large w limit), so that the equation becomes the wave equation

$$\frac{\partial^2 g}{\partial z^2} = \frac{\partial^2 g}{\partial w^2}$$
(43)

One can solve this immediately by the method of characteristics. The characteristics are $w+z =$ a constant, and $w-z =$ a constant, straight lines in wz space. Let us say that g(w) is the boundary condition at $z=z_o$, and we are interested only in solutions for $z < z_o$. Calling the known $f_1$ at $z=z_o$, $f_1(w)$, we know that $g(w) = w^{-(Z+1)}f_1(w)$

Then the solution for all w and $z<z_o$ is which is equal to g(w) at $z=z_o$ is

$g(w,z) = ag(w+z_o-z) + (1-a)g(w-z_o + z)$   (44)

Here a is a constant. We know that at large w, g(w) has to approach zero in order to be a physically reasonable distribution function. Hence the as z decreases, the solution must decrease, that is the solution has to have the electrons lose energy as they move toward smaller z. The second term has the electrons accelerating as they go to smaller z. This means that only the first term is a physically reasonable solution, and the proper characteristic solution is

$g(w,z) = g(w+z_o-z)$             (455)

However the solution also has to be zero for all w at $z =0$. In other words, there must be a 'reflected wave' starting at the origin, canceling out the 'incident wave' and propagating back toward positive z. Hence the solution accounting for this is

$$g(w,z) = g(w+z_o-z) - g(w+z_o+z) \quad (46),$$

and

$$f_1(w,z) = \left[\frac{w}{w+z_o-z}\right]^{Z+1} f_1(w+z_o-z)$$
$$- \left[\frac{w}{w+z_o+z}\right]^{Z+1} f_1(w+z_o+z)$$

$$(47)$$

Thus one has a simple formula for $f_1$. Not only that, it does not depend on a sparse eigenfunction expansion being possible. Of course in the particular case, one has to examine where in wz space the neglect of the last term in Eq(42) is valid. It is almost never valid everywhere in that space.

In any case to get a very rough first approximation for the preheat, one might use Eq (47) as a first approximation for the distribution function. Once one has this, one finds the relative energy flux of the energetic electrons by using Eq. (9).

It is also noteworthy that if a system is described simply by Eq.(46), and $g(w,z_o)$ is simply a summation of a few exponentials like $\exp-(k_n w)$, then the solution by the method of characteristics, and the solution by the method of sparse eigenfunctions, give *exactly* the same solution for $g(w,z)$. This gives additional credibility to the use of the sparse eigenfunction method.

Appendix B: Reflexing electrons

To continue, we discuss a point made in (4), namely that a sizable fraction of the energetic electrons produced may never make it into the target, because they are reflexing around the outer part of the blowoff plasma. These cannot give rise to heating of either the ablator or fuel, because they never make it in. Here is where the difference in attitude between Ref (4) and this paper becomes significant. Reference (4) discusses the experiment as a completed process, so it produces so many reflexing electrons, period. This work discusses the instantaneous *rate* of production of energetic electrons, reflexing or not, and correspondingly, their energy *flux*. However the reflexing electrons cannot simply accumulate, they must ultimate go somewhere, and if they do not reenter the target, another electron from somewhere else must reenter the target to prevent charge buildup in the target.

Figure (18) envisions several possibilities.

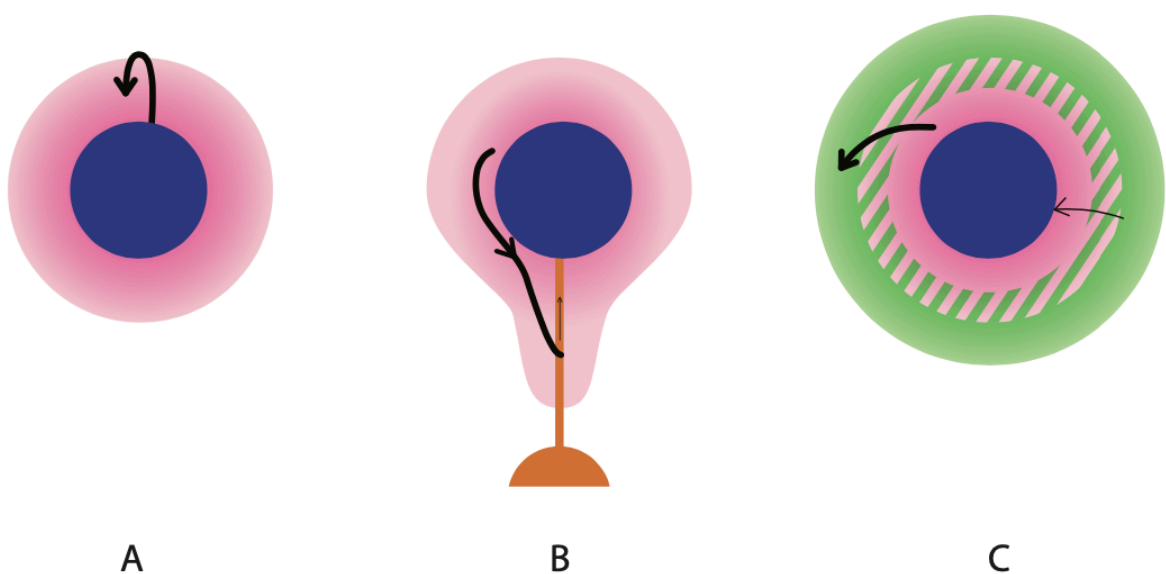

A  B  C

Figure (18): Several possible configurations involving energetic electrons. The blue is the target, pink is the region of energetic electrons, orange is a target stand, green is a plasma around the target, a heavy black line is a typical energetic electron path, a light black line the path of a slow electron which enters the target; A: The target is in a vacuum and is not supported in any way. No matter how many times an energetic electron reflexes around the target, it ultimate enters it, since it has nowhere else to go without charging the target. B: The target is mounted on a stand shown. Some of the reflexing electrons may go there and low energy may return to the to the target through the stand. C: The pink and green regions intersperse. The the target is not supported, but is in a gas which the energetic electrons partially ionize. Then an original energetic electron could simply disappears into the gas plasma, and a low energy electron from the plasma returns to the target to prevent it from charging.

Most likely B is the configuration of the NIF/URLLE experiment. However in a laser fusion power plant, most likely, the targets will be shot in at high speed in a vacuum or near vacuum, as pictured in A. Clearly it would be much more complicated to send in targets on a succession of stands, and it is unlikely that the situation will be viable if there is a background pressure. This background would certainly be turbulent, considering the energy release in the chamber, and it is not clear that the target would ever reach the focal spot of all the lasers. Certainly the issue of tracking the target on the way in is much more difficult if the target travels in through a turbulent gas. Hence assuming the configuration for a laser fusion power plant is A, all energetic electrons produced by the instability will ultimate strike the target whether or not they reflex around it a few times. If that is the case, the NIF/URLLE experiments may be giving an overly optimistic view of the the effect of the instability on laser fusion, by not counting many of the electrons which, in a laser fusion situation, will ultimately reenter the target and give rise to preheat, but in the current experiment configuration, likely do not.

Appendix C: The distribution function just beyond the quarter critical surface

The distribution as shown in Figs (5 and 6) from the simulation shows a distribution function very much localized in angle. Clearly a two term Legendre expansion is not a very good approximation to it,. However the simulation (14) only gives the distribution function right up to, and slightly beyond, the quarter critical density. The size of the simulation region is in fact not even a mean free path for the electrons, so it cannot show the transformation of the distribution function from the straight line orbits, Fig. ( 9A) to a more random orbits of Fig. (9B).

Clearly we must have a good approximation of the distribution function a bit further into the denser plasma. These energetic electrons scatter and they are attempting to become isotropic. However they cannot become completely isotropic a short distance into the denser plasma, because then they would have no energy flux, whereas the distribution function shown in Figs (5 and 6) clearly has a significant energy flux into the target. If the distribution function really had no energy flux just inside the quarter critical surface, that is if it were isotropic, then the large drop in energy flux would mean there would be an enormous heating of the plasma just inside the quarter critical surface. Clearly this is unphysical, especially since in the electron ion scattering, the electron energy is conserved.

Hence the distribution function just inside the quarter critical surface must be as isotropic as possible, but also must have the same temperature, and energy flux as that shown in Figs (5 and 6) from the simulation, and as implied from the experiment. Recall that the experiment did not measure the energy flux, but we assumed that the ratio of energy flux to laser energy flux is the same as the ratio of energetic electron energy to total laser energy.

Hence, we assume that just inside the quarter critical surface, the distribution function $f_1$ has the w dependence of a displaced Maxwellian with the same temperature and energy flux as the experiment implies, and/or, the simulation shows. That is for $f(w,\theta,z_o)$ we take:

$$f(w,\theta,z_o) = \exp-(u^2 - 2uu_o\cos\theta + u_o^2) \qquad (48)$$

Here $u = w^{1/4}$, and $u_o$ corresponds to the displacement of the assumed Maxwellian in the direction of the spatial variation (i.e. in the z direction for planar, and in the r direction for spherical). The energy flux of the instability generated electrons comes only from the $u_o$ term. Hence $u_o$ is determined so that the energy flux of the distribution function in Eq. (10) is equal to the calculated or measured energy flux of the energetic electrons at the quarter critical surface.

Since the assumed velocity of the thermal energy flux is assumed to be much less than thermal velocity of the energetic electrons, we know that $u_o \ll 1$. The distribution function can be very simply expanded in a series of series of Legendre polynomials of $\cos\theta$, not simple a power series expansion in $\cos\theta$. However for the first two terms, the expansion is the same. Beyond that there are odd higher order terms in the $\cos\theta$ expansion which contribute to the first order term in the Legendre expansion.  However these corrections are all small by various powers of $u_o$.

Hence we approximate

$$f_1(w, z_o) = 2uu_o \exp{-u^2} = \alpha u \exp{-u^2} \quad (49)$$

where $\alpha$ is determined by the energy flux at $z = z_o$ and to get the complete distribution function, we must multiply by $\cos\theta$.

Acknowledgement:  This work was supported by DoE NNSA and ONR.  The author thanks Steven Obenschain, the head of the NRL Laser fusion group for supporting this work even though the author had only minimal contact with the group due to the pandemic.  This previous contact has proven to be invaluable and the author hopes to not only keep it up, but to greatly increase it.  He especially thanks both Dr Michael Rosenberg, and especially Dr.  Andrey Solodov, both of URLLE,  for help interpreting the data that URLLE presented in Ref. (4).  He also thanks  Dr. C. Xiao of the Key Laboratory for Micro-/Nano-Optoelectronic Devices of Ministry of Education, School of Physics and Electronics, Hunan University, Changsha, 410082, China for help interpreting his simulations and also for providing Fig. (6), data which did not appear in their Ref (13).  This work initially began in a cooperative effort by the author with Denis Colombant.  However due to circumstances, Dr. Colombant was not able to continue.  The author very much appreciates the help he was able to give here, and especially to the cooperative efforts of Refs. (1,2,3 and 14).